\documentclass[10pt,aps,prx,floats,twocolumn,amsfonts,amssymb,unsortedaddress,floatfix]{revtex4}

\usepackage{epsfig}
\usepackage{amssymb}
\usepackage{mathtools}
\usepackage{color}
\usepackage{upgreek}

\usepackage{caption}

\newcommand{\quotes}[1]{``#1''}

\usepackage{siunitx}

\usepackage{tikz}
\usepackage{esint}

\usepackage{hyperref}
\hypersetup{
    colorlinks,
    citecolor=black,
    filecolor=black,
    linkcolor=black,
    urlcolor=black
}

\begin{document}
\title{Emergent dimer-model topological order and quasi-particle excitations\\ in liquid crystals: combinatorial vortex lattices
}

\author{$^1$Cuiling Meng, $^1$Jin-Sheng Wu, $^{2,3,4,5}$Žiga Kos,  $^{2,3}$Jörn Dunkel, $^{6*}$Cristiano Nisoli, $^{1,2,7,8*}$Ivan I. Smalyukh}

\affiliation{$^1$Department of Physics and Chemical Physics Program, University of Colorado, Boulder, CO 80309, USA}
\affiliation{$^2$International Institute for Sustainability with Knotted Chiral Meta Matter (WPI-SKCM$^2$), Hiroshima~University, Higashihiroshima, Hiroshima, Japan}

\affiliation{$^3$Department of Mathematics, Massachusetts Institute of Technology, Cambridge, MA 02139, USA}

\affiliation{$^4$
 Faculty of Mathematics and Physics, University of Ljubljana, Ljubljana, Slovenia
}
\affiliation{$^5$
Department of Condensed Matter Physics, Jožef Stefan Institute, Ljubljana, Slovenia
}

\affiliation{$^6$Theoretical Division, Physics of Condensed Matter and Complex Systems, and Center For Nonlinear Studies, Los Alamos National Laboratory, Los Alamos, NM 87545, USA}
\affiliation{$^7$Department of Electrical, Computer, and Energy Engineering, Materials Science and Engineering Program, University of Colorado, Boulder, CO 80309, USA}
\affiliation{$^8$Renewable and Sustainable Energy Institute, National Renewable Energy Laboratory, University of Colorado, Boulder, CO 80309, USA}
\affiliation{$^*$Corresponding authors: cristiano@lanl.gov and ivan.smalyukh@colorado.edu}

\date{\today}

\begin{abstract}

Liquid crystals have proven to provide a versatile experimental and theoretical platform for studying topological objects such as vortices, skyrmions, and hopfions. In parallel, in hard condensed matter physics, the concept of topological phases and topological order has been introduced in the context of spin liquids to investigate emergent phenomena like quantum Hall effects and high-temperature superconductivity. Here, we bridge these two seemingly disparate perspectives on topology in physics. Combining experiments and simulations, we show how topological defects in liquid crystals can be used as versatile building blocks to create complex, highly degenerate topological phases, which we refer to as \lq Combinatorial Vortex Lattices\rq~ (CVLs). CVLs exhibit extensive residual entropy and support locally stable quasi-particle excitations in the form of charge-conserving topological monopoles, which can act as mobile information carriers and be linked via Dirac strings. CLVs can be rewritten and reconfigured on demand, endowed with various symmetries, and modified through laser-induced topological surgery — an essential capability for information storage and retrieval. We demonstrate experimentally the realization, stability, and precise optical manipulation of CVLs, thus opening new avenues for understanding and technologically exploiting higher-hierarchy topology in liquid crystals and other ordered media.

\end{abstract}

\maketitle

\section{Introduction}

Topology provides a unifying framework for classifying fundamental structural properties across a wide range of physical systems~\cite{topophys,knotphys, smalyukh2020knots}. Yet, applications of topological concepts in different areas of physics can have quite different flavors.  Practitioners in the fields of fluid dynamics, liquid crystals (LCs) and elasticity, have long and successfully been using topology to classify field configurations that cannot be morphed one to another by a continuum deformation when described by different topological invariants~\cite{softmatterphys}. Another notion of {\it topological order} has emerged to explain high-temperature superconductivity~\cite{anderson1987resonating, moessner2010quantum} and the quantum Hall effect~\cite{laughlin1983anomalous}, with applications to  spin liquids~\cite{wen1989vacuum, sachdev2018topological, wen2002quantum} and in quantum computation~\cite{nayak2008non}. Such quantum topological order, and its classical analog~\cite{henley2010coulomb,henley2011classical}, pertains to the space of all possible low-energy degenerate or quasi-degenerate configurations in discrete many-particle systems consisting of two (or more)  species. The number of such low-energy configurations can grow combinatorially with the system size, and they can often be partitioned into homotopy classes called topological sectors~\cite{macdonald2011classical,castelnovo2012spin,jaubert2013topological,nisoli2020topological,zhang2023topological}. Although these configurations typically appear highly disordered, they obey topological laws -- a situation described as \lq constrained disorder\rq. Thence, violations of the topological constraints on the disorder manifest themselves as charge-conserving fractionalized excitations that can only be created and annihilated in pairs of opposite charge~\cite{kivelson1987topology, moessner2001resonating}.
Because of their inherent robustness, these topological quasi-particles are promising candidates for novel information storage and computation paradigms~\cite{di2022memcomputing,gartside2022reconfigurable}.

\begin{figure*}[tb!]
  \includegraphics[width=2\columnwidth]{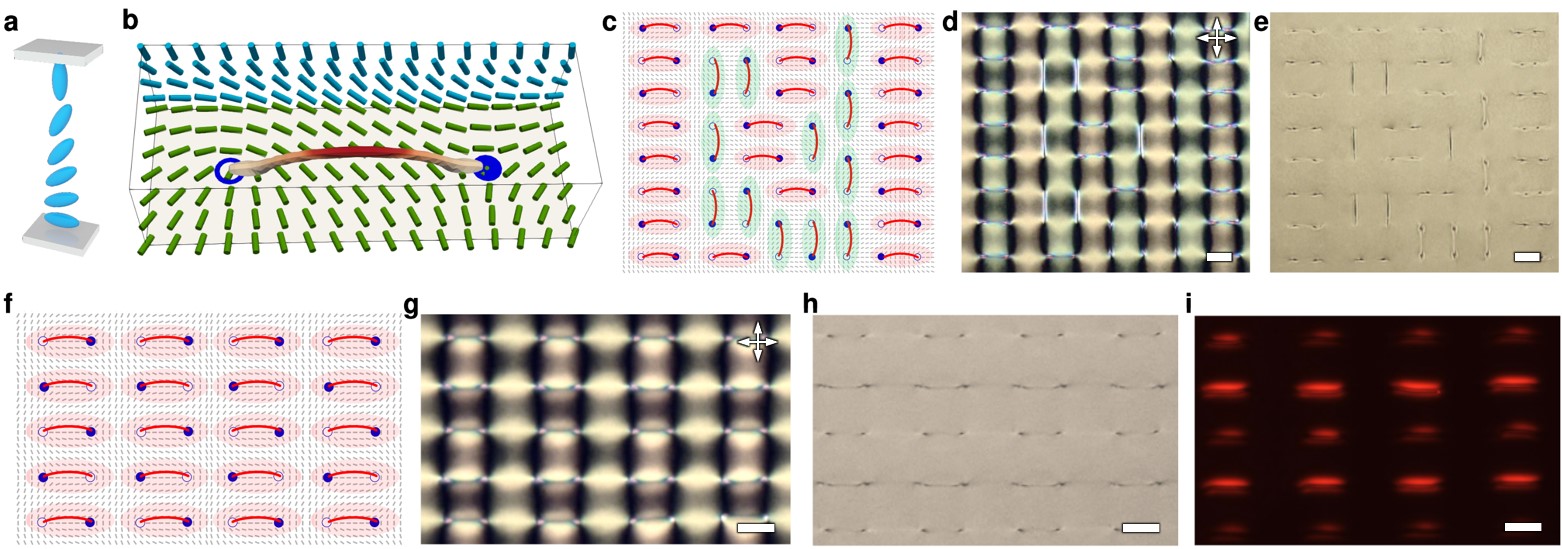}
  \caption{
  {\bf Dimer model experimentally realized in liquid crystal-based}  combinatorial vortex lattices.}
  {\bf a,}~Vortex lattices are created  in LC cells with patterned tangential director orientation at the bottom surface and uniformly perpendicular director orientation at the top surface.
  {\bf b,}~Using photopatterning, we prescribe the director profile at the bottom surface. This panel shows a numerical simulation with a patterned texture of a $+1/2$ and $-1/2$ nematic singularities at the surfaces connected by a vortex line. Such texture generates a vortex line, pinned to the surface and extending into the bulk~\cite{PetitGarrido}.
  {\bf c--e,}~We pattern a square lattice of $+1/2$ and $-1/2$ pinning sites and observe a set of vortex lines connecting pairs of pinning sites. Each vortex line is described as a single dimer. The sketched dimer arrangement in (\textbf{c}) corresponds to (\textbf{d}) the polarizing optical microscopy and (\textbf{e}) bright-field microscopy, respectively.
  {\bf f--i,}~Pre-engineered aligned state with horizontal dimer orientations as schematically visualized in (\textbf{f}) is experimentally observed under polarized optical microscopy (\textbf{g}),  bright-field microscopy (\textbf{h}), and phase-contrast microscopy (\textbf{i}), respectively. Thin gray rods in (\textbf{c},\textbf{f}) depict the photopatterned surface boundary conditions at the bottom surface. Double white arrows with black frames mark the orientations of the crossed polarizers. All scale bars are \SI{50}{\micro\metre}.

  \label{fig:1}
\end{figure*}

\begin{figure*}[t]
  \includegraphics[width=1.8\columnwidth]{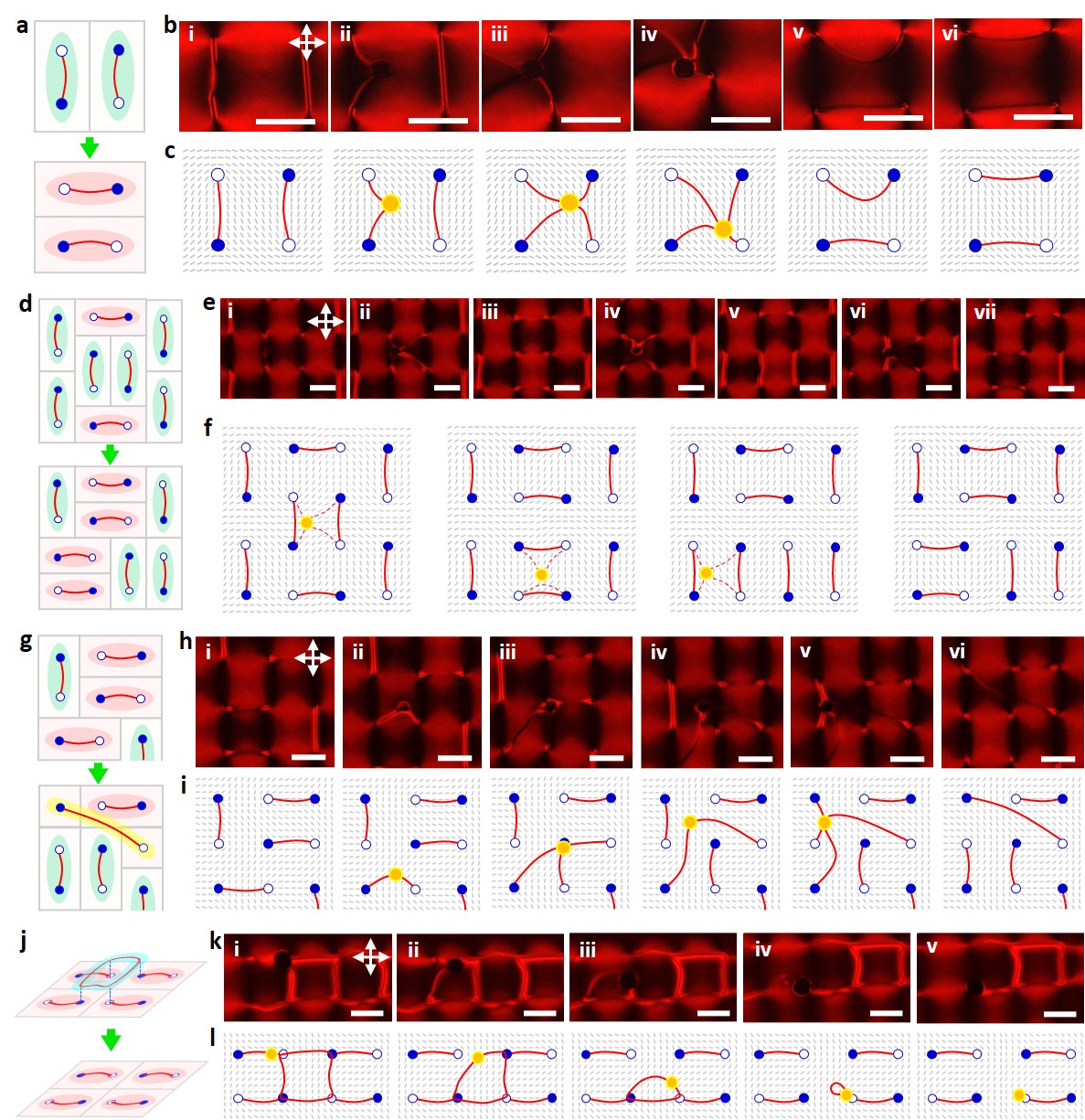}
  \caption{ 
  \textbf{Topological surgery on dimer arrangements.} 
  \textbf{a,}~Elementary update of dimers by flipping a pair of vertex lines. 
  \textbf{b,c,}~Snapshots and corresponding schematics showing how the optical tweezer slowly drags one vortex line to approach, merge and then re-connect with the other. Gray rods map the LC director field $\mathbf{n}(\mathbf{r})$ as surface boundary conditions at the bottom surface and yellow filled circle marks the position of the tweezer. \textbf{d,}~Collective renewal of dimer arrangements. 
  \textbf{e,f,}~Sequence of snapshots and schematics visualizing the initial, transitional, and eventual dimer ensembles. The red dotted lines in the first three panels in (\textbf{f}) correspond to the rewiring motion in panels ii, iv and vi in (\textbf{e}). 
  \textbf{g,}~Growing of dimers distinguished by a yellow-highlighted line represents an elongated and diagonally wiring dimer. 
  \textbf{h,i,}~Snapshots and corresponding schematics showing that the tweezer moves slowly to overcome increasing elastic energy and wires the vertex lines beyond nearest neighbor lines. 
  \textbf{j,}~Annihilation of loop-profile dimers (blue-highlighted line). 
  \textbf{k,l,}~Experimental snapshots and schematics showing the moving trajectory of tweezer and the accompanying collapse of the closed-loop vertex line. Double white arrows mark the orientations of the crossed polarizers. All scale bars are \SI{100}{\micro\metre}. 
  }
  \label{fig:2}
\end{figure*}

\begin{figure*}[tb!]
\renewcommand{\figurename}{}
  \includegraphics[width=1.8\columnwidth]{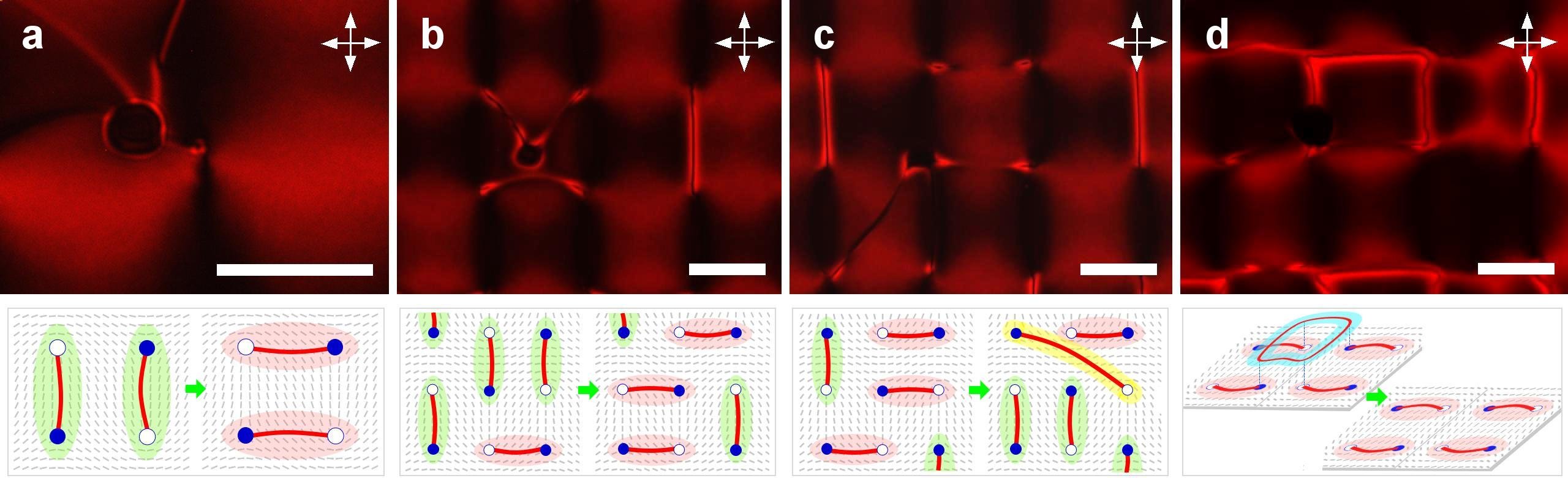}
  \caption*{ 
\textbf{\href{https://www.overleaf.com/project/651fd68e0478f68fa56bba26}{VIDEO 1.} Topological surgery enabled by the optical tweezer.} 
  \textbf{a-d.}~Polarized optical microscopy textures showing elementary update of dimers, collective renewal of dimers, “growing” of dimers and annihilation of a loop-profile dimer (blue highlighted shade), respectively. Sketched dimer arrangements before and after the optical surgery are provided below. Double white arrows mark orientations of crossed polarizers. The videos are recorded by fixing the optical tweezer while moving the LC cells in x-y plane. Scale bars are \SI{100}{\micro\metre}.  
}
  \label{VIDEO:1}
\end{figure*}

Since the detection of topological phases in natural systems has proved quite challenging, recent research has focused on deliberately realizing these phases in artificial systems. Examples include  nanomagnets~\cite{perrin2016extensive,lao2018classical,zhang2023topological}, colloids~\cite{libal2018ice}, and microwave guides~\cite{chen2019experimental}, or even mechanical metamaterials~\cite{sirote2024emergent}. These engineered platforms facilitate the study of fundamental and exotic phenomena such as magnetic monopoles, ergodicity breaking, exotic memory effects~\cite{perrin2016extensive,nisoli2020topological,lao2018classical,zhang2023topological}, topological charge transfer~\cite{libal2018ice}, topologically-constrained kinetics ~\cite{king2021qubit,lopez2023kagome}, nonabelian manipulation~\cite{sirote2024emergent}, classical analogues to topological entanglement entropy~\cite{chen2019experimental}, and skyrmions~\cite{duzgun2021skyrmion,tai2023field}. However, there are currently no reconfigurable experimental platforms for implementing and controlling topological order in soft condensed matter systems.

Here, we close this gap by introducing experimental realizations of \lq Combinatorial Vortex Lattices\rq~ (CVLs) in LCs. CVLs are assemblies of topological defects that are anchored to pinning sites and strategically arranged (Fig.~\ref{fig:1}) to realize a combinatorial manifold of disordered configurations. The extensively degenerate ensemble of configurations of these CVLs mirrors and generalizes the topological models studied in quantum and classical statistical physics~\cite{kasteleyn1961statistics,Baxter1982,anderson1987resonating,moessner2010quantum,kivelson1987topology}. 

Because the configuration space of CVLs grows exponentially with system size, it yields a residual entropy comparable to that of frustrated magnets, with implication for information encoding and neuromorphic computations. Importantly, CVLs support charge-conserving quasi-particle excitations. Note that these are higher-hierarchy topological defects, distinct from the usual nematic defects of LCs. In other words, we employ conventional topological defects in liquid crystals as building blocks to construct an higher-level topology that is, instead, unconventional in LCs.

Compared to previous artificial realizations of classical topological order, our CVLs offer unparalleled versatility---they are designable, fully controllable, re-writable, and experimentally characterizable, as we now show. 

\section{Combinatorial Vortex Lattices} 
 
We experimentally realize rewritable CVLs by creating lattices of nematic vortex lines within an LC medium where the vortex lines connect photopatterned substrates at lattice vertices \cite{Martinez,meng2023topological}. In this arrangement, vortex lines serve as dimers in a two-dimensional dimer model~\cite{kasteleyn1961statistics,kenyonintroduction}. A dimer model involves dimers covering lattice edges, ensuring each node has precisely one and only one dimer. Our CVL-based realization of the dimer-model corresponds  to the prototypical system of classical topological order, employed first in the context of high $T_c$ superconductivity~\cite{anderson1987resonating} and widely studied ever since. Other relevant (and more general) models can also be realized as CVLs, as we will show later.  

Vortex lines in LCs are singular, topologically protected lines of discontinuity for the average orientation of rodlike molecular axes, described by the nonpolar LC director field $\mathbf{n}(\mathbf{r})$~\cite{smalyukh2020knots}. In our experiments and simulations, the LC is confined within a cell between top and bottom glass plates of distinct boundary conditions (Fig. 1a). One plate establishes translationally invariant molecular orientations perpendicular to the substrates, while the other plate controls tangential-to-surface orientations of LC molecules to form pre-engineered spatial patterns (depicted respectively by the blue and green rods in Fig. 1b). In our experiments, these patterns are coordinated in the form of optically controlled molecular orientations within a 10-nm thin layer of azobenzene dye on the glass surface \cite{Martinez,meng2023topological,meng2018hybridcell}. The dye orientations are in practice controlled by linearly polarized blue-light illumination, causing its long molecular axes to orient orthogonally to the linear polarization. By synchronizing the blue-light illumination area with the corresponding light polarization using a half-wave waveplate and polarizer, we can engineer the azo-dye orientation profile with a pair of $\pm 1/2$ topological defects (Fig. 1c-i). Each unit within the $\pm 1/2$ lattice contains a vortex core, due to a symmetry breaking effect, as the orientation of LC molecules is ill-defined in the vortex center. The corresponding $\pm 1/2$ lattice textures, along with a group of vortex lines, are clearly observed in the polarized optical micrographs of Fig. 1d and 1g. Notably, when compared to bright-field microscopy (Fig. 1e,1h), the visibility of vortex lines or dimers is enhanced under phase-contrast microscopy (Fig. 1i).

Figure 1 illustrates the realization of a nematic dimer model on a square lattice in which vortex lines play the role of dimers. Figure~\ref{fig:1}c and 1f show a disordered and ordered dimer cover configuration respectively. The number of possible configurations can be computed exactly and scales exponentially with the lattice size, yielding an extensive entropy with promising utility in information encoding. Specifically, the entropy per dimer is~\cite{kasteleyn1961statistics}  $s=2G/\pi$ where $G$ is the Catalan constant, $G\simeq 0.916$.

\begin{figure*}[t]
  \includegraphics[width=\textwidth]{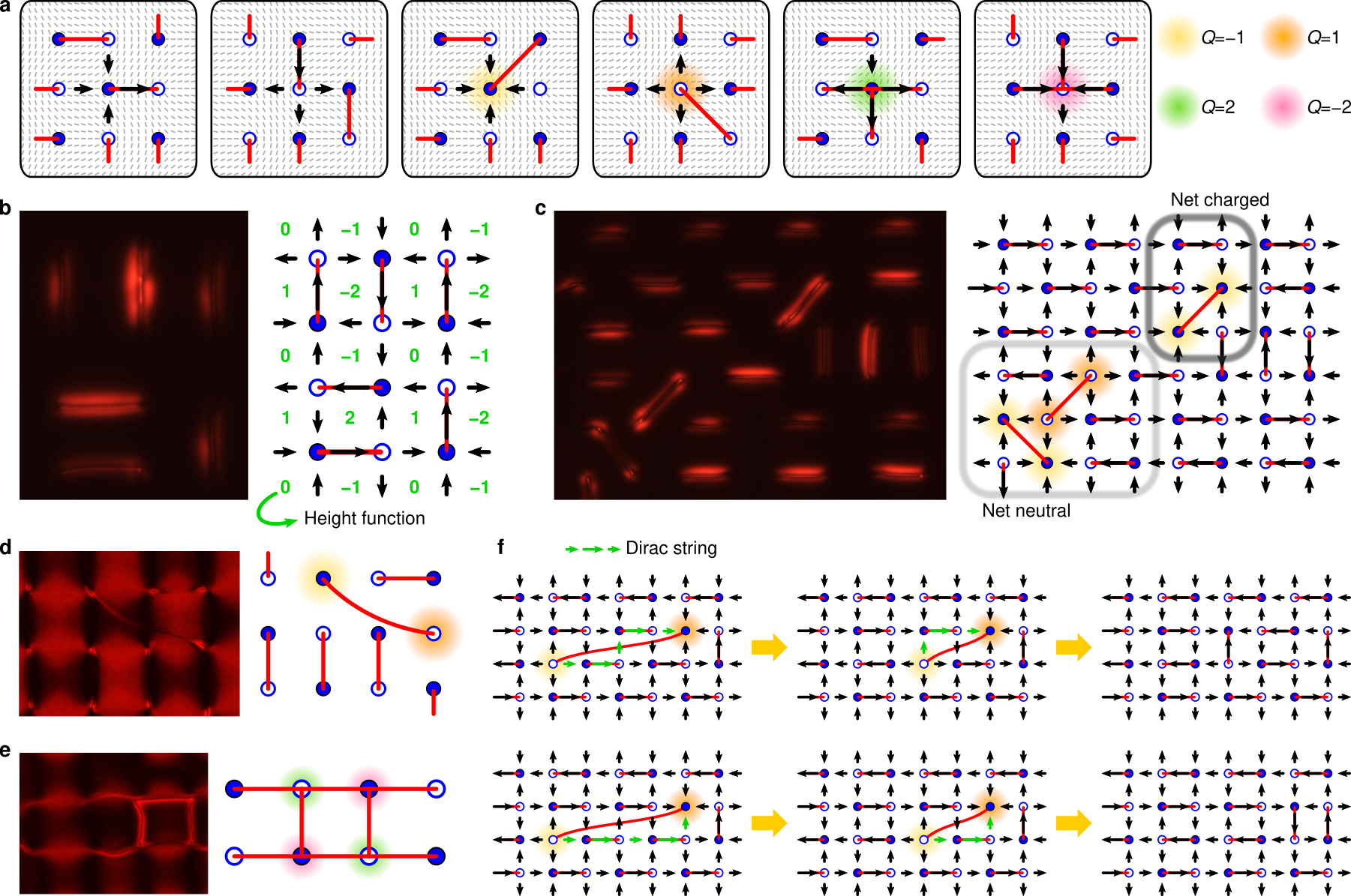}
  \caption{
  {\bf Quasi-particle excitations and Dirac strings in LC dimer lattices.} 
  {\bf a,}~A dimer configuration can be expressed as a pseudo-magnetic field (black arrows on the edges of the square lattice). The blue pinning site has inwards flux along the non-dimerized edges and {three-times larger} outwards flux along the dimerized edge. The white pinning site has outwards flux along non-dimerized edges and {three-times larger}  inwards flux along dimerized edge. The divergence of the field around a pinning site represents its pseudo-charge $Q$. The net charge of an improperly dimerized pinning site is non-zero, corresponding to quasi-particle excitations. 
  {\bf b,}~A properly dimerized lattice generates a divergence-free pseudo-magnetic field that can be described by a height function $h$ as $\nabla\times(0,0,h)$.
  {\bf c,}~Example of a dimer ensemble with defects in the form of diagonal dimerization, obtained by laser tweezer rewiring. The diagonal dimer in the dark gray box cannot be eliminated by line-reconnection with the nearby dimers, because the net charge within the box is $-2$; instead the two diagonal dimers in the light gray box can, because their net charge is 0.
  {\bf d,}~Two monopoles with $Q=\pm 1$ can be spatially separated, where their energetic costs scales approximately linearly with their mutual distance.
  {\bf e,}~Vortex loop consisting of triple-dimerized pinning sites can be annihilated in experiments (see Fig.~\ref{fig:2}) since it has a zero net charge.
  {\bf f,}~The annihilation trajectory of a monopole pair along the Dirac strings represented by green arrows connecting $+1$ and $-1$ monopole. Two rows show two possible choices (among many) for the Dirac string and the subsequent rewiring steps. All scale bars are \SI{100}{\micro\metre}.
    }
  
  \label{fig:3}
\end{figure*}

\section{Direct manipulation with laser tweezers} 

Our CVLs are rewritable and can be precisely manipulated via  optical tweezers, to either reconfigure the dimer cover ensemble or also create hierary chically higher-order topological defects as charged quasi-particles.

The topological nature of a dimer configuration is apparent in that it can only be changed by a collective rearrangement involving multiple dimers~\cite{oakes2016emergence}. The simplest local update that conforms to the covering rules involves the flipping of only two parallel dimers~\cite{henley1997relaxation}. 
Figures 2a-2c show how the elementary update of a configuration can be performed by using optical tweezers to obtain a 
guided re-connection of vortex lines, a form of “topological surgery” analogous to chromosome meiosis~\cite{antoniou2017extending}. 

Technically speaking, for  optical tweezers operating at a wavelength of 1064 nm, the infrared beam spot heats up the conductive layers on the inner surfaces of two glass plates. The generated heat subsequently transfers to the thermotropic LC causing it to transition from nematic to isotropic states as indicated by a black spot in panels ii-iv in Fig. 2b. Following this transition, the optical gradient and elastic forces associated with the moving tweezers allow us to drag or elongate the vortex line (Fig. 2b, 2c and panel a in VIDEO 1). 

An iteration of this elementary update enables a remaking of the dimer cover, as  Fig.~\ref{fig:2}d-2f show (see also panel b in VIDEO 1). And yet, importantly,  optically guided reconnections can also generate topological defects at a hierarchically higher level. Notice how in Fig. 2g-2i and panel c in VIDEO 1, for instance, each step of a ``growing" topological surgery, accomplished through a series of reconnections among non-parallel dimers, results in vortex lines that do not lie on the edges of the square lattice. The resulting configurations are therefore not proper dimer covers, as indicated by the long, yellow-highlighted vortex line in the schematics of Fig. 2g. The non-parallel, long dimers can reversibly undergo ``shrinkage", ``re-connection" and ``extra growing" with the use of optical tweezer and deliberate manipulation. 

\begin{figure*}[ht!]
  \includegraphics[width=2\columnwidth]{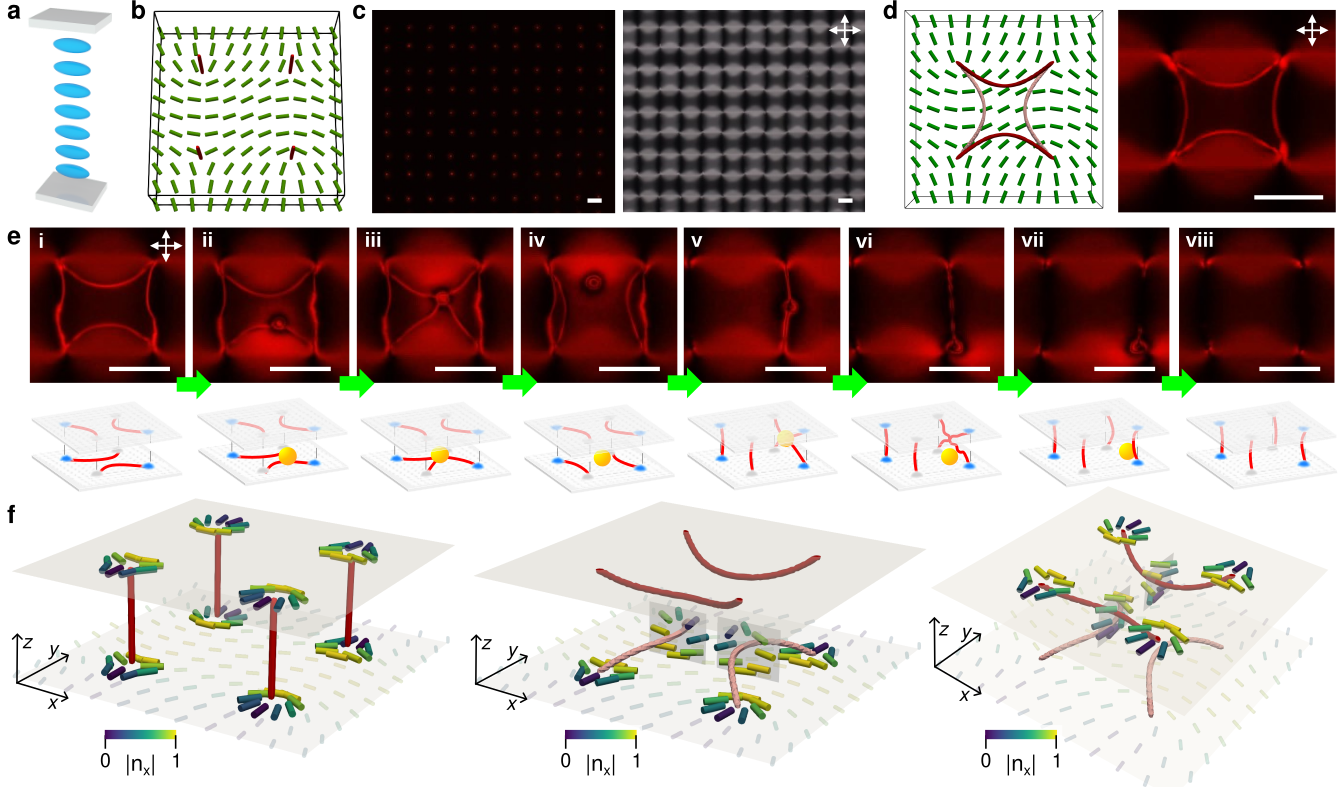}
  \caption{
  {\bf Three-dimensional LC dimer models.}
    {\bf a,b,}~By photo-patterning $\pm 1/2$  lattice of singularities at both the bottom and top confining surfaces, we can create vortex lines vertically connecting the pinning sites through the depth of the cell. 
    {\bf c,}~Phase-contrast microscopy (left) and corresponding polarized optical microscopic texture (right) of such vertical dimers.
    {\bf d,}~Simulated and experimental polarized optical image showing three-dimensional vortex lines with in-plane wiring on the top surface (dark red) and the bottom surface (light red).} 
    {\bf e,}~Polarized microscopic snapshots (top) and corresponding schematic (bottom) showing the evolution of dimers from horizontal to longitudinal wirings. Double white arrows mark the orientations of the crossed polarizers.
    {\bf f,} Three-dimensional views of panels (b,d) showing the director field structure at selected vortex line cross-sections.
    All scale bars are \SI{50}{\micro\metre}. 
    
    \label{fig:4}
\end{figure*}

Lastly, other types of defects can be introduced, such as ``overdimerizations" where the pinning sites anchor more than one vortex line. In the situation  illustrated in Fig. 2,j-l (and panel d in VIDEO 1), a loop of dimers has been added to a regular dimer cover (Fig. 2j) so that each node is dimerized three times. Then, Fig.~2k and Fig.~2l demonstrate how the loop can be removed through successive reconnections.

\section{Pseudo-magnetic field, charged quasi-particles, and Dirac strings} 

We can now characterize these defects, in particular to assess whether they can be removed through local manipulations. Local stability is important if defects are to be employed as  mobile information carriers.

Figure~\ref{fig:3}a shows that a CVL configuration can be described by an {\it emergent pseudo-magnetic field}: each edge of the square lattice is assigned an arrow such that, if an edge is unoccupied by a dimer, the arrow has a length of 1, pointing toward a \quotes{blue} vertex and away from a \quotes{white} vertex. Conversely, if the edge is occupied, the arrow points in the opposite direction with a length of 3. Then, the net flux of the pseudo-magnetic field around any closed loop is zero by construction, and in two dimension, one can introduce a pseudo-magnetic potential (the so-called ``height function") whose rotated gradient corresponds to the pseudo-magnetic field (Fig.~\ref{fig:3}b). Intuitively, the fluctuations of this height function label the disorder of the dimer phase. This formalism allows for the identification of defects as charged quasi-particles. The charge is equal to the phase of the height function, or equivalently to one fourth of the flux of the pseudo magnetic-field, as shown in Figs.~\ref{fig:3}a, which accounts for all the various defects previously introduced in Fig.~\ref{fig:2}. Note that in literature an equivalent picture is presented, based on an {\it emergent pseudo-magnetic field}~\cite{henley2011classical,lao2018classical}. The two pictures are equivalent(in two-dimension), by rotating the arrows by 90 degrees, and replacing the flux of the field with its circulation.

\begin{figure*}[tb!]
  \includegraphics[width=1.8\columnwidth]{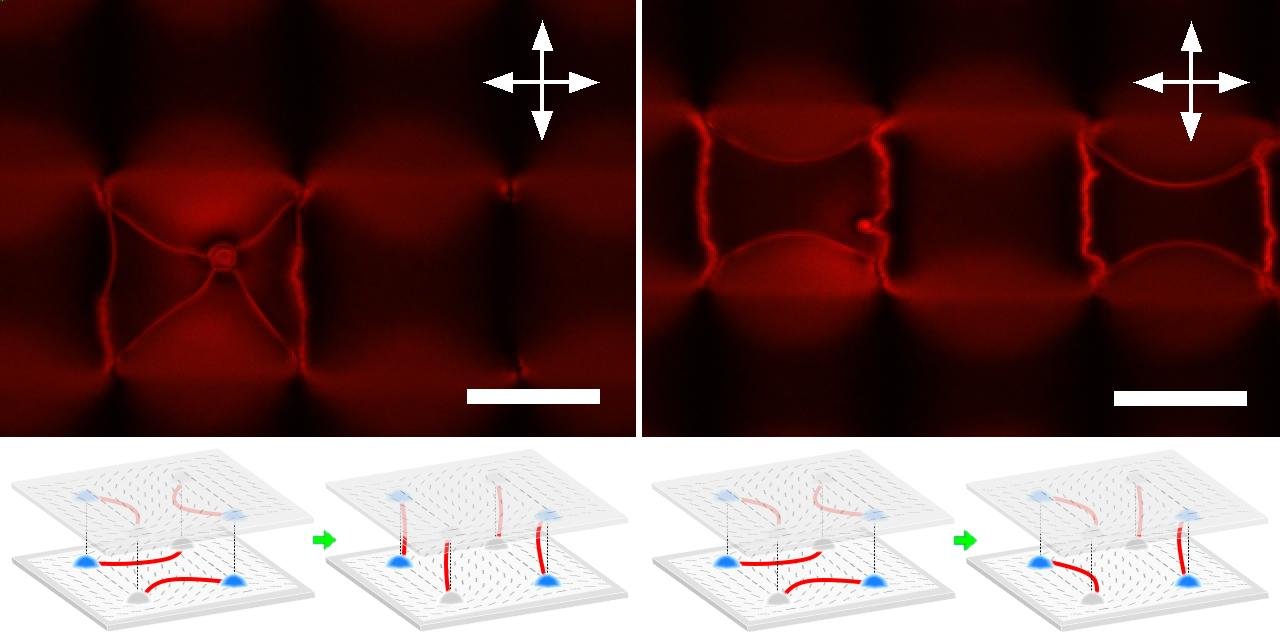}
  \caption*{ 
\textbf{\href{https://www.overleaf.com/project/651fd68e0478f68fa56bba26}{VIDEO 2.} Renewal of dimer configurations in a three-dimensional fashion.} Transformation from in-plane vertical dimers into complete longitudinal ones (left) and into partially longitudinal ones (right). Corresponding three-dimensional sketched diagram are provided below. The movie was obtained by fixing the laser spot while mobilizing the sample between fixed crossed polarizers (double white arrows). Scale bars are \SI{50}{\micro\metre}. 
}
\end{figure*}

\begin{figure*}[t]
  \includegraphics[width=1.8\columnwidth]{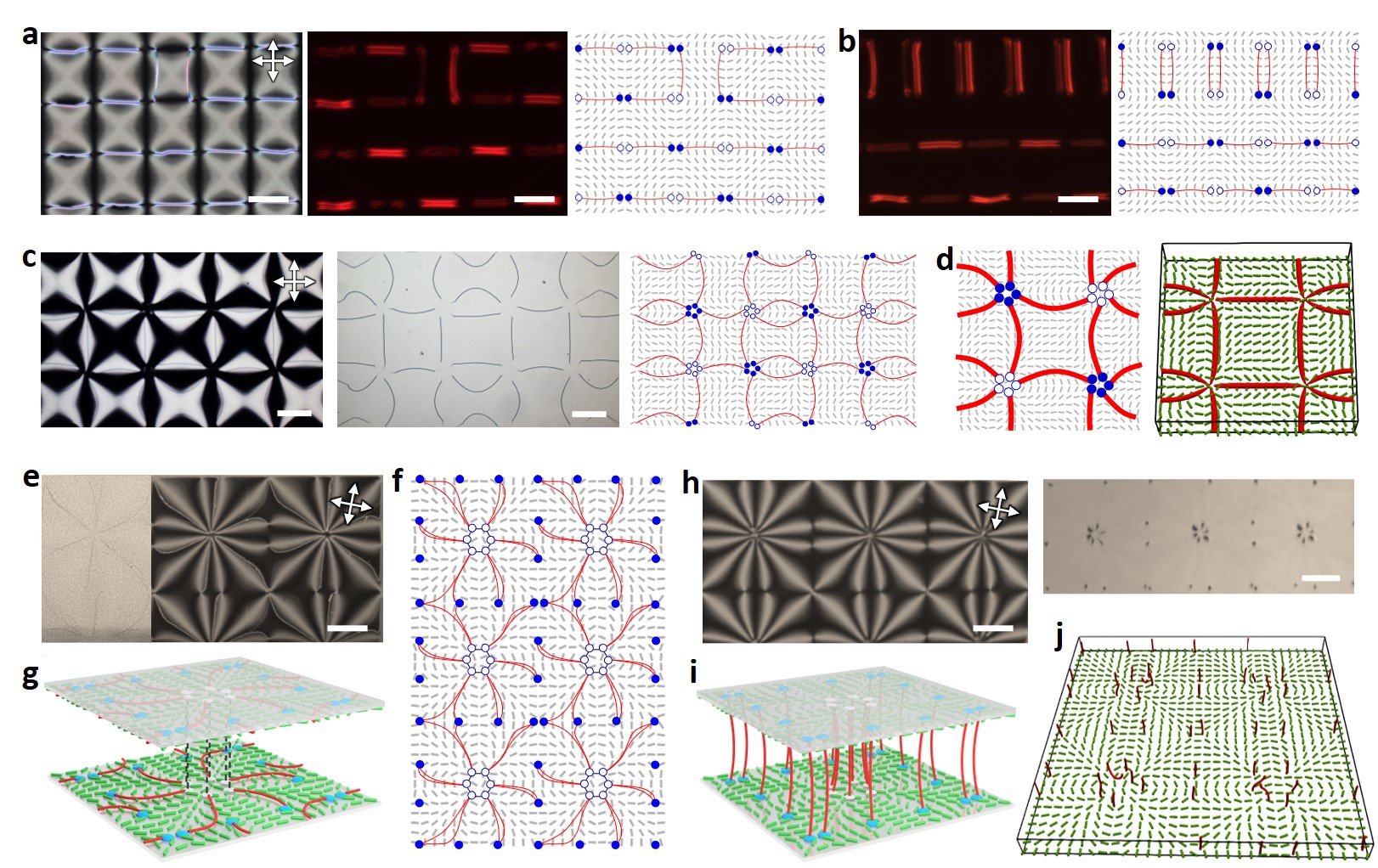}
  \caption{
  {\bf  Realization of higher-dimer models.}
  {\bf a,}~A double-dimer ensemble in a $\pm 1$  lattice landscape characterized by polarized optical image (left) and phase-contrast micrograph (middle) and schematic drawing (right), respectively. 
  {\bf b,}~Reconfiguring on the double-dimer in (\textbf{a}) by flipping dimers.
  {\bf c,}~Five-dimer model realized by $\pm 5/2$  lattice. The core of each 5/2 defect is energetically split into five pinning sites as observed from the microscopic images with (left) and without (middle) crossed polarizers and as visualized by the schematic picture (right). {\bf d,}~Zoom-in schematic of a repeating unit (left) and corresponding simulated three-dimensional structure (right).
  {\bf e,}~Six-dimer model based on +3 square lattice in a planar LC cell. Each +3 singularity is encircled by twelve alternating bright and dark brushes as viewed from its polarized micrograph and splits into 6 vortex lines on each surface as shown. 
  {\bf f,g,} Schematic drawings of a 2×3 lattice in two-dimension ({\bf f}) and a single unit in three-dimension ({\bf g}).
  {\bf h,}~The horizontal lines in ({\bf e,g,f}) are optically transformed into longitudinal ones as demonstrated by corresponding micrographs captured with and without crossed polarizers. 
  {\bf i,j,}~The corresponding three-dimensional schematic of a single unit and simulated structure of a 2×2 sub-lattice. Double white arrows with black frames mark the orientations of the crossed polarizers. All scale bars are \SI{200}{\micro\metre}.
  }
  \label{fig:5}
\end{figure*}

Crucially, the net charge of a cluster of defects tells us its stability. Fig.~\ref{fig:3}c showcases various defects, captured through phase-contrast microscopy, together with a schematic representation.  Consider the four charges in the light-grey rectangle in Fig.~\ref{fig:3}c: they correspond to two charges $=+1$ and two charges $=-1$: because their total charge is zero,  they can be eliminated by performing two reconnections among vortex lines. Now consider instead the diagonal dimer in the dark-grey rectangle (Fig. 3c). 
No reconnection with its neighbors can eliminate this defect because the total charge inside the dark-grey rectangle amounts to $-1-1=-2\ne 0$. Reconnections merely shift the location of the charge pair. To eliminate the diagonal vortex line we need to move it next to another one of charge  $1+1=2$.
Similarly, the monopole pair in Fig.~\ref{fig:3}d is charge neutral and it could be annihilated, but also even further separated. The over-dimerized vortex loop in Fig.~\ref{fig:3}e consists of two $Q=-2$ and two $Q=+2$ charges. Its charge-neutrality means that it can be annihilated through local rewirings as already shown in Fig. 2j-2l.

Finally, charge-preserving defects in topological phases are associated with a {\it Dirac string} --  akin to flux lines connecting magnetic monopoles in spin ice~\cite{Castelnovo2008}. For example, consider Fig.~\ref{fig:3}f, which shows two opposite charges $\pm 1$ connected by a vortex line, and the emergent pseudo-magnetic vector field. Then, a line of vectors all pointing head-to-toe represents the flux of a possible Dirac string -- and thus also a path for the mutual annihilation of the two defects, as depicted in Fig. 3f. Interestingly, in this case the defects connected by a Dirac string are linearly confined, as the energy of the vortex line scales with its length: thus the two charges feel an attraction (see the discussion at the end of Ref.~\cite{nisoli2020topological}).

\section{Going to the third dimension} 

Our CVL methodology can be extended to the third dimension with a dimer model on a cubic lattice realizing a so-called bi-layer Coulomb phase~\cite{desai2021bilayer}.
 When the $\pm 1/2$ lattice is assembled in an LC cell with tangential surface boundary conditions (Fig. 4a), where the inner surfaces of both glass plates are photo-patterned, each defect core accommodates two pinning sites independently pinned on the top and the bottom surfaces. Consequently, within a repeat unit containing two $1/2$  and two $-1/2$ singularities on each surface, it is possible for the lines to wire in a three-dimensional fashion by either wiring + and $-1/2$ pinning sites at lattice vertices on each inner surfaces inseparably (Fig. 4d and panels i-iv in Fig. 4e) or wiring $+\,+$ or $–\,–1/2$ pinning sites across the space of two inner surfaces (Fig. 4b and panels vii-viii in Fig. 4e). The LC director field $\mathbf{n}(\mathbf{r})$ of each + or $-1/2$ surface singularity in the lattice is revealed by two alternatively dark and bright brushes as shown in the polarized optical micrograph (Fig. 4d). 
 However, the LC molecules within the lines are distorted giving rise to LC symmetry-breaking on the core. The lines, therefore, scatter imaging red light and appear to be a red-dot array as viewed from the top surface of $\pm1/2$ lattice by phase-contrast microscopy (Fig. 4c). 
 
In this new scenario, we can still reconfigure the lines using optical tweezers, but with greater freedom due to the additional space dimension. This includes adjustments both within planes and across the lines in the plane, relative to those vertically across the LC layer, as illustrated by the snapshots in Fig.~4e. Real-time manipulation is shown in VIDEO 2. Conversely, vertical lines can be also dragged by the tweezer and reconnected into lines floating in parallel to the glass plates. Note that the dimer ensemble in panel viii of Fig.4e is in a more stable configuration as the length of vortex lines ($\sim$ 10--30 \SI{}{\micro\metre}) are one order of magnitude lower than that in panel i of Fig. 4e ($\sim$ 100--200 \SI{}{\micro\metre}). That is because the system’s elastic energy is proportional to the length of the lines. 

\section{Beyond conventional dimer models--future directions}

We have so far focused on the dimer model, as a particularly simple yet paradigmatic topological phase completed with quasi-particle excitations. However, the scope of realizations possible within CVLs is remarkably vast and will require further exploration in future work. We conclude this report by demonstrating the potential of realizing novel models with intriguing and still unexplored properties, such as multiple dimer systems \cite{kenyon2014conformal}.
By introducing multiple pinning sites into the lattice nodes, we can generate new combinatorial configurations. Specifically, placing two pinning sites yields what is known as a double-dimer model. 

This model comprises two interlinked replicas of the standard dimer model (Fig. 5a,5b), is currently studied by mathematicians~\cite{kenyon2010double,kenyon2011boundary} and involves scaling and conformal invariance \cite{kenyon2014conformal}.
More generally, the number of pinning sites at vertices is determined by the winding number $k$ of a topological singularity placed at the vertices of photopatterned arrays, and is twofold of $k$. This is driven by minimization of the free energy of vortices that extend from surfaces into the nematic bulk, which scales quadratically with $k$ and tends to replace one vortex with 2$k$ half-integer counterparts. For example, each vortex in ±1 lattice yields two pinning sites (Fig. 5a,5b), and similarly, each vortex in ±5/2 lattice is occupied by five pinning sites (Figs. 5c-5e,6), etc. 

Hence, the addition of more pinning sites to the nodes of a pre-designed lattice allows for a natural extension of the concept of a double-dimer model to a multi-dimer model. Specifically, by constructing a +3 lattice within the three-dimensional landscape of the nematic bulk confined by two photopatterned plates, we can manipulate a super dimer ensemble comprising six-, double-, and single-dimer models within a single cell (Fig. 5f-5k).

Here, multiple $-1/2$ and $-1$ surface singularities surrounding each +3 singularity serve to compensate for the +3 charge and meet the requirement of a zero net charge. The number of $-1/2$ vortex connecting each vertex of the lattice can vary, either being larger or smaller than that of the near-neighbour vertices. This variation can be controlled to be uniform throughout the lattice or to exhibit spatial variations. Moreover, the configuration is reconfigurable optically through re-patterning or using laser tweezers, or electrically by applying voltage to confining electrodes~\cite{meng2023topological}. This demonstrates the rich variety of models that can be realized using our method.

\section{Discussion and outlook} 

Advancing recent interdisciplinary efforts to implement topological models~\cite{lieb1972inphase, Baxter1982,henley2010coulomb,henley2011classical,lao2018classical} into artificial materials~\cite{perrin2016extensive,lao2018classical,zhang2023topological,libal2018ice,sirote2024emergent,king2021qubit,lopez2023kagome}, we have introduced Combinatorial Vortex Lattices (CVLs) as a versatile experimental platform for realizing both prototypical as well novel types of dimer models in nematic LCs.  CVLs employ pinned nematic defects as building blocks to realize topological phases and quasiparticles of conserved charge that are locally stable and enable robust information encoding. We showed how CVLs can be reconfigured using laser surgery, and that quasi-particle charge manipulation can be interpreted as driven mobility of stable information carriers. 
Similar to artificial spin ice systems, which are also often described by dimer models~\cite{Baxter1982,macdonald2011classical,lao2018classical},   CVLs present an ensemble of collectively constrained bits that can provide a testbed for  novel information encoding and computation paradigms~\cite{di2022memcomputing,gartside2022reconfigurable,caravelli2020logical}.

\par
More broadly, CVLs hold promise for studying thermal~\cite{pivsljar2022blue} or electromagnetically driven kinetics, by allowing the direct optical observation of fundamental phenomena such as slow relaxation, topologically constrained kinetics, and ergodicity breaking.  Furthermore, since the energetics of CVLs can be finely tuned by using materials with different elastic constants or by applying external electromagnetic fields~\cite{ackerman2017diversity,meng2023topological},  the design space for effective CVL Hamiltonians is vast. For example, future implementations could incorporate models of interacting dimersin which the ground-state degeneracy is replaced with a quasi-degeneracy of proliferating metastable states~\cite{kim1994critical,alet2005interacting,wilkins2020interacting}. Such models have been of interest in the study of glassiness for computational purposes~\cite{barahona1982computational}.
\par
A particularly appealing aspect of CVLs, both from a scientific and a technological perspective, is their high degree of controllability and tunability. While we focused above on laser manipulation, nematic LCs are highly responsive to electric and magnetic fields ~\cite{FerroSwitching,meng2023topological}, enabling the integration of multiple sequential or parallel control mechanisms. 
Specifically,  recent advances in dynamic non-contact photo-alignment technology allow for spatial patterning and re-configuration of complex LC textures ~\cite{wei2014optvortices,nys2022photoaligned,jiang2024moire,mcginty2021large,Neyts2018PAtoVA,gorkunov2020LCmetasurfaces,Martinez} and open the possibility of optically writing and reading information in CLVs. To enhance the information density, multi-beam-based approaches offer a path for constructing CVLs at submicron scales ~\cite{neyts2024multiimpose,Zhi2023lightpulse,Egor2013nano}. CVL miniaturization promises a broad range of future applications, including multilevel anti-counterfeiting and identification, topological photonics, electro-optics, reconfigurable diffractive optics, solitonic light localization \cite{meng2023topological}, colloidal assembly \cite{Martinez} and other technologies exploiting the multiplicity of states in CVLs.
\par
Scientifically, the ability to create complex CVL  designs on demand and to control the topological invariants that characterize individual vortices by realizing CVLs in LC media exhibiting different symmetries opens vast opportunities for fundamental research. Our present study focused exclusively on demonstrating the rich phenomenology of CVLs in uniaxial nematic LCs.  An exciting future challenge will be the creation of even more complex CVLs in nematic media with lower symmetry, such as orthorhombic and monoclinic biaxial nematics, ferromagnetic colloidal nematics, or chiral LCs~\cite{wu2024natcom,mundoor2021nat,mundoor2018sci,liu2016pnas}, which possess different groundstate manifolds and, therefore,  exhibit topologically different types of vortex line defects. For example, the non-Abelian vortex lines corresponding to elements of the quaternion group emerge in biaxial nematic LCs \cite{mundoor2018sci}, and chiral nematic systems and are expected to exhibit nontrivial connectivity at the vertices of CVLs \cite{wu2024arxiv}. Exploring and understanding the combinatorial diversity of such topological super-structures will require concerted experimental and theoretical efforts over the next decade.

\newpage

\appendix

\begin{figure*}[t!!!]
  \includegraphics[width=\textwidth]{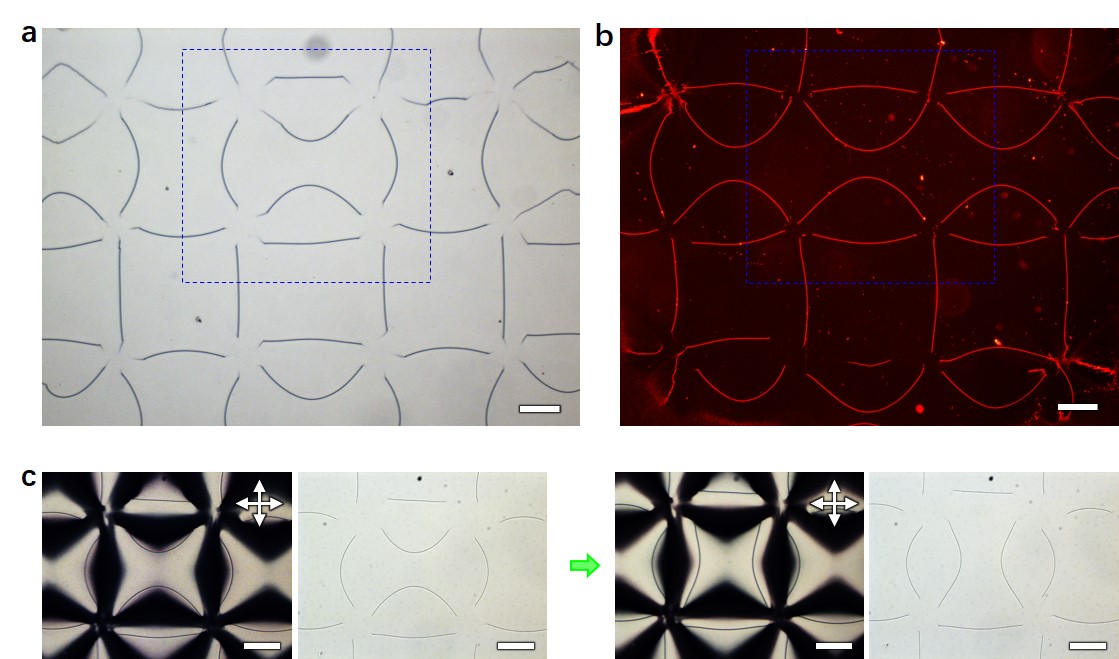}
  \caption{\textbf{Five-dimer model in a square lattice of $\pm 5/2$ vortices.} Five pinning sites on each vortex core allow five dimers to extend and pin with neighboring sites to architect a five-dimer model. Lines arrangement in (\textbf{a}) can be translated to that in (\textbf{b}) by repeatedly flipping the dimers in pairs as shown in (\textbf{c}). The vortex lines appear as dark and bright red under bright-field microscopy (\textbf{a}) and phase-contrast microscopy (\textbf{b}), respectively. \textbf{c,} The polarized (left) and bright-field (right) images before and after a pair of dimers flipping. Double white arrows with black frames mark the orientations of the crossed polarizers. All scale bars are \SI{100}{\micro\metre}. 
  }
\end{figure*}

\section{Sample preparation} 

The LC cells were made with two types of glass slides or coverslips. Glass plates of type I was treated with 5.0 wt\% polyimide SE5661 (Nissan Chemicals) by spin-coating at 2,700 rpm for 30 s and then baking at 90°C, followed by 1 h at 180°C to set strong perpendicular boundary conditions for $\mathbf{n}(\mathbf{r})$ at the LC-glass interface. Glass plates of type II were spin-coated using 1.0~wt\% azobenzene dye SD1 in dimethylformamide at 3,000~rpm for 45~s and subsequently baked on a hot stage at 100$^\circ$C for 10~min to evaporate the residual solvent. To build the glass cells for the LC confinement, glass plates of type I and type II were assembled into a hybrid-mode cell as schematized in Fig. 1a. Alternatively, two glass plates of type II were sandwiched into a planar-mode cell as schematically shown in Fig. 4a. Specifically, a fast-setting epoxy glue containing silica spacer spheres (with diameters ranging from 10 to \SI{30}{\micro\metre}, from Thermo Fisher) was placed near the corners of one glass surface, the other glass plate was then lapped atop. After epoxy solidified, the glass cells were then photo-patterned with a predefined geometry for $\mathbf{n}(\mathbf{r})$ to induce the desired single topological vortex or their lattices via the method described in the following. After the photo-patterning, the cells were filled with a nematic LC, 4-cyano-4'-pentylbiphenyl (5CB; EM Chemicals) using capillary forces. To avoid unintentional effects of ambient light on azobenzene dye alignment layers, the samples were stored under dark-room conditions.

\textbf{}
\begin{figure*}[tb!]
  \includegraphics[width=\textwidth]{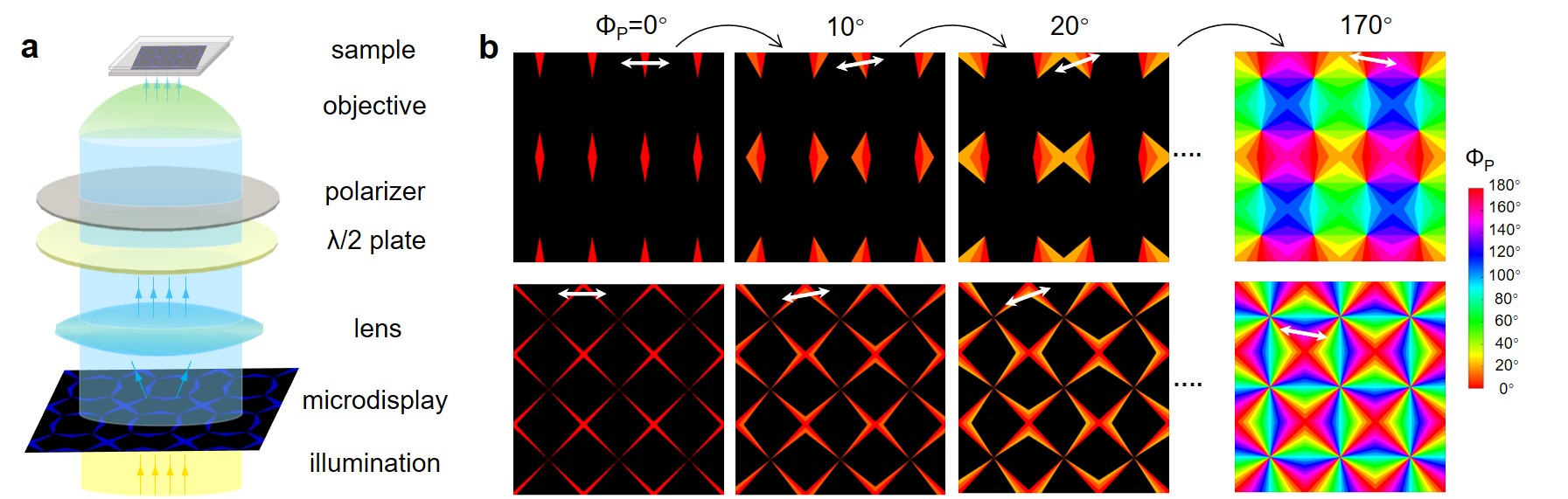}
  \caption{\textbf{Principles of photo-patterning for pre-defined vortices array based on azobenzene dye. }\textbf{a}, Schematic of a custom-built setup for spatial photo-patterning arbitrary geometry on the azobenzene layers. A microdisplay with dense pixels is encoded with a pre-designed blue-light pattern. The pattern that initially diverges after the microdisplay is converged and collimated after a set of lenses. The polarization of the collimated blue light is set to be a desired direction by rotating a half-wave plate and a linear polarizer. The polarized blue light impinges the back aperture of the objective and finely focuses on the azobenzene dye layer on the inner surfaces of confining plates. The azobenzene dye is blue-light sensitive and energetically favorable to rotate its long axes into a direction orthogonal to the light polarization. \textbf{b}\textbf{,} Vortices lattices such as $\pm 1/2$ (top row) and $\pm 2$ (bottom) arrays are accomplished via sequential illuminating the segments separably, and each segment is set with certain linear polarization (double white arrows) to photo-align the azobenzene dye at corresponding tangential direction (${\phi}_P$). After successive segments refreshing and polarization tunning, a complete pattern encoded with vortices geomotry is made. LC molecules, after infiltrating into the cell, are templated by the patterned azobenzene dye and form the desired lattice.     
 }
\end{figure*}

 \textbf{}
  \begin{figure*}[tb!]
    \includegraphics[width=\textwidth]{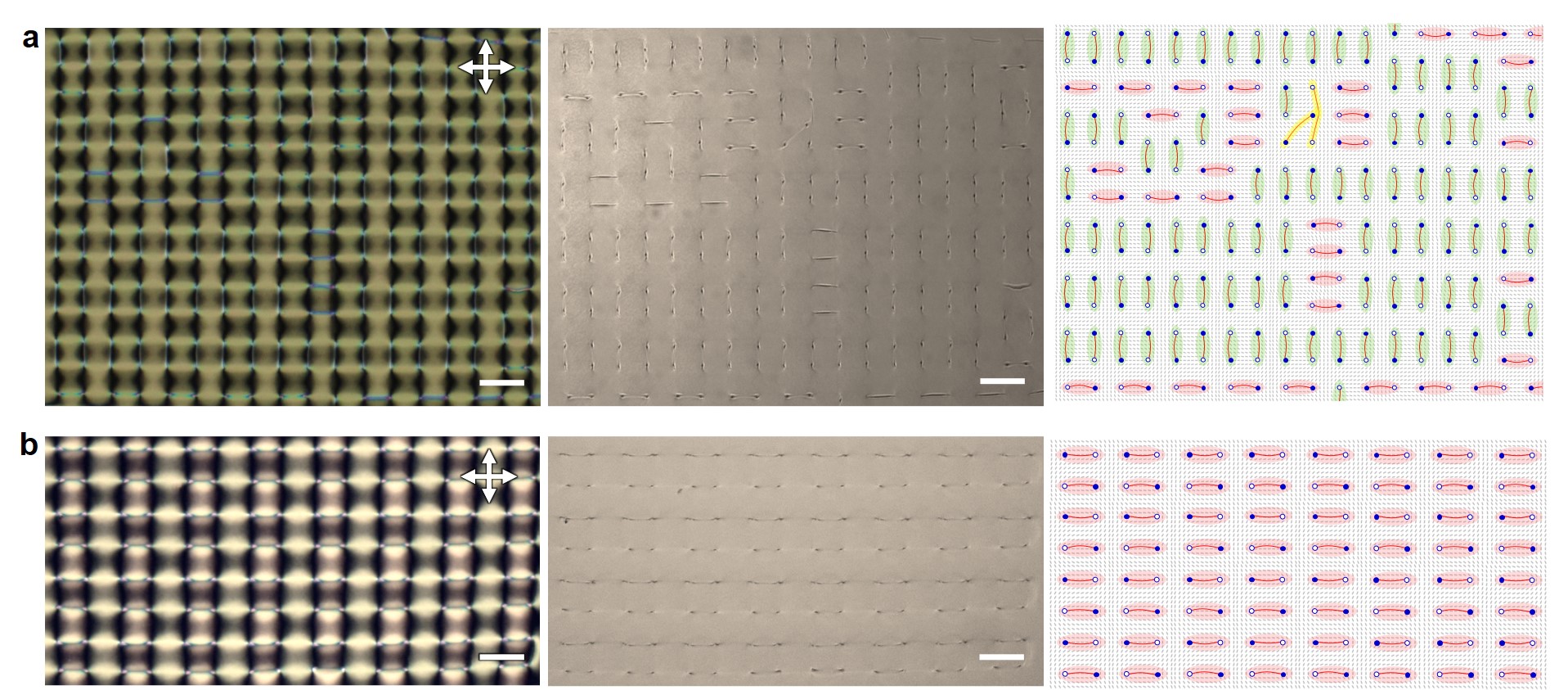}
  \caption{\textbf{Dimer model with a larger scale $\pm 1/2$ vortices lattice compared to that in Fig. 1.} \textbf{a,b,} Dimer model with constrained disorder (\textbf{a}) and orderly aligned configuration (\textbf{b}). Panels from left to right in (\textbf{a}) and (\textbf{b}) are polarized optical micrographs, bright-field micrographs, and corresponding sketched dimer arrangements, respectively. Double white arrows with black frames mark the orientations of the crossed polarizers. All scale bars are \SI{100}{\micro\metre}. }
\end{figure*}

\section{Design of vortices lattice}

Square periodic arrays of vortices were designed as surface boundary conditions of LC-filled volumes confined between glass plates~\cite{meng2023topological}.
We first created patterns of two-dimensional vortices to define the spatially varying boundary conditions before arranging them into various periodic arrays.
In each defect, considering the energetic stability of vortices~\cite{tsung2021crystal}, we adopted the ansatz for nematic director $\mathbf{n}=\cos(\varphi)\hat{x}+\sin(\varphi)\hat{y}$ with the angle between the director and $\hat{x}$-axis being $\varphi=k\phi+\varphi_0$, where $k$ gives the topological charge (integers of 1/2), $\phi$ is azimuthal angle measured from the vortex point, and $\varphi_0$ is a phase shift that defines local spatial rotations of the nematic director field by the same value without changing the topological charge.

The fabrication of vortex arrays could be done simply by spatially placing the two-dimensional units of the arrays (Figs.~5e-5j,6) and then relaxing the director field while minimizing the free energy. For instance, the lattices with $k=3$ implemented with this procedure have additional defects generated spontaneously due to the topological constraint of zero net charge~\cite{tsung2021crystal}. In other cases, the pre-defined vortices and their topology-compensating counterparts (with opposite charges) were placed alternatingly into square-periodic or other arrays that also have zero net charges. As suggested by the geometry, the values of $\varphi_0$ were chosen such that the boundary is smooth for vortex lattices, with $k$ being half integers.

\section{Photo-patterning of vortices lattice }

Our mesostructured patterns of vortex lines were built by spatially connecting field configurations corresponding to the arrays of vortex-pinning sites. Each pinning site was generated by light-controlled orienting of azobenzene dye molecules to define boundary conditions corresponding to each nematic vortex with a desired winding number. To locally anchor  LC molecules and director in a specific direction at the LC-glass interface, the inner surfaces of the according confining glass plates were functionalized (coated) with an $\sim$10-nm-thin layer of the azobenzene dye, a polarization-sensitive light-responsive photoalignment material. Upon being illuminated with a small dose ( $\sim$200~mJ~cm\textsuperscript{–2}) of linearly polarized blue light, the azobenzene moieties of the dye molecules tend to orient orthogonally to the polarization direction of the illuminating light, yielding robust spatially controlled surface boundary conditions~\cite{meng2018hybridcell} \textbf{n}(\textbf{r}) for the LC (Figs. 1,4 and 5). By using a commercial  LC microdisplay, all desired two-dimensional boundary condition geometries for\textbf{ n}(\textbf{r}) can be predefined through spatially illuminating linearly polarized blue-light patterns, as controlled on a pixel-by-pixel basis with computer software. As schematically shown in Fig. 7a, a blue-light pattern generated by a computer is projected through the microdisplay with 1,024~×~768~pixels (EMP-730, Epson) and then relayed by a lens module to the back aperture of an objective. The blue light imprinting the pattern is focused on the azobenzene dye layers. For the planar-mode cells, the two azobenzene dye layers on the opposite inner surfaces of the confining glass plates can be patterned for boundary conditions sequentially or at the same time, depending on the cell gap relative to the focus depth of the objective~\cite{subhash2011full}. In this study, we did simultaneous patterning of the two surfaces of the cells by using low-magnification ×4 objectives (numerical aperture of 0.13) and relatively small cell gaps. The desired linear polarization was controlled by a half-wave plate and a linear polarizer inserted at appropriate locations along the optical path (Fig. 7a). By means of multi-step illuminations synchronized with the linear polarization rotation, any pre-designed structure of boundary conditions can be generated at will.

To define LC vortices and their lattices, as schematically illustrated in Fig. 7b, the entire intended structure was split into angular segments of roughly constant in-plane director orientation, with the number of segments determining the smoothness of photopatterned structures. The \textbf{n}(\textbf{r}) boundary condition for each pattern of the same angular segment were generated by the same linearly polarized excitation light projected onto the sample. After multiple sequential projections done for different azimuthal orientations of \textbf{n}(\textbf{r}), a complete geometry of lattice of two-dimensional vortices was imprinted into the spatially oriented azobenzene dye layer. The angular resolution of patterning, defined by discrete changes of the linear polarization direction between two adjacent segments, was controlled to be within 1° to 45°, depending on the need. For example, 10° angular resolution utilized in the photo-patterning of ±1/2 defect lattices gives a rather smooth texture revealed by its polarized optical microscopy (Figs. 1d,1g and 4c-4e). Lateral dimensions of the light-programmed \textbf{n}(\textbf{r}) structures and lattice parameters can be varied from micrometers to millimeters, as defined by the projected size of the blue-light pattern and the magnification of an objective. Examples are shown in Fig. 8, which provides a large-scale visualization of disordered and ordered dimer cover configurations emerging as LC vortex lattices upon filling in LC into glass cells.

The dimer patterns spontaneously emerge once a cell with patterned confining substrates is filled with the nematic LC, albeit it can be reconfigured by melting the LC within the entire cell or locally by using laser tweezers. From the LC perspective, the vortex lines are energetically costly and typically tend to take the shortest distance between two pinning sites. The multiplicity of different connections between the pinning sites can be probed by quenching such samples from isotropic state many times. On the other hand, once the system is in the nematic phase, to change the vortex orientation one needs to provide enough energy for a vortex line to be extended and recombined with another vortex line. Such topological surgery was performed using an optical tweezer system that locally melts the nematic order and allows one to extend the vortex line. When this process joins two vortex segments, they can reconnect and rewire into a new configuration. In our experiments, once a dimer pattern is spontaneously generated after the cell being filled with the nematic LC, laser tweezers allow us to reconfigure it into desired dimer patterns by laser-guided reconnection using the optical tweezer system.

\section{Optical imaging and control with laser tweezers}

Olympus optical microscopes working in transmission and reflection geometries were utilized in the reported studies.  A charge-coupled device camera (Grasshopper3, from PointGrey) mounted on an upright BX-51 or inverted IX-71 microscope (both from Olympus) was used for optical video microscopy. Olympus objectives with ×2, ×4 and ×10 magnifications and numerical apertures within 0.06–0.4 were used for bright-field optical imaging. The microscopes were additionally equipped with pairs of insertable and rotatable linear polarizers and both quarter-wave and half-wave plates, which we utilized for the polarizing optical microscopy experiments. 

The laser tweezers set-up, which we used to optically manipulate the LC vortex lines, is based on an ytterbium-doped fibre laser (YLR-10-1064, IPG Photonics, operating at 1,064~nm) and a phase-only spatial light modulator (P512-1064, Boulder Nonlinear Systems) with 512~pixels~×~512~pixels, each with a size of 15~$\upmu$m~×~15~$\upmu$m~\cite{ackerman2017diversity,RTweezerPiestun,martinez2014mutually} integrated with an inverted IX-81 optical microscope (Olympus)~\cite{smalyukh2010three}. The beam from the laser is first reflected off the spatial light modulator and then projected to the back aperture of an objective. The computer-generated holograms are supplied to the spatial light modulator by computer software at a rate of 15~Hz, ensuring real-time pre-envisaged manipulation within the plane (Figs. 2,9) and across the LC cells (Figs. 10,11). In either cases, the net charge of a cluster of defects should be zero to ensure the reconnection among vortex lines as shown in Figs. 3c and 12. Polarized optical images revealing details of laser manipulation are captured and videos are recorded using the same Grasshopper3 charge-coupled device camera. At a nominal laser power of ~1~W, a ×20 objective with a numerical aperture of 0.5 was used to focus the laser beam for locally heating and melting the LC by prompting a transition from the nematic to isotropic phase and then quenching back to the nematic state.

\section{Numerical simulations}

\begin{figure*}[th!!]
  \includegraphics[width=\textwidth]{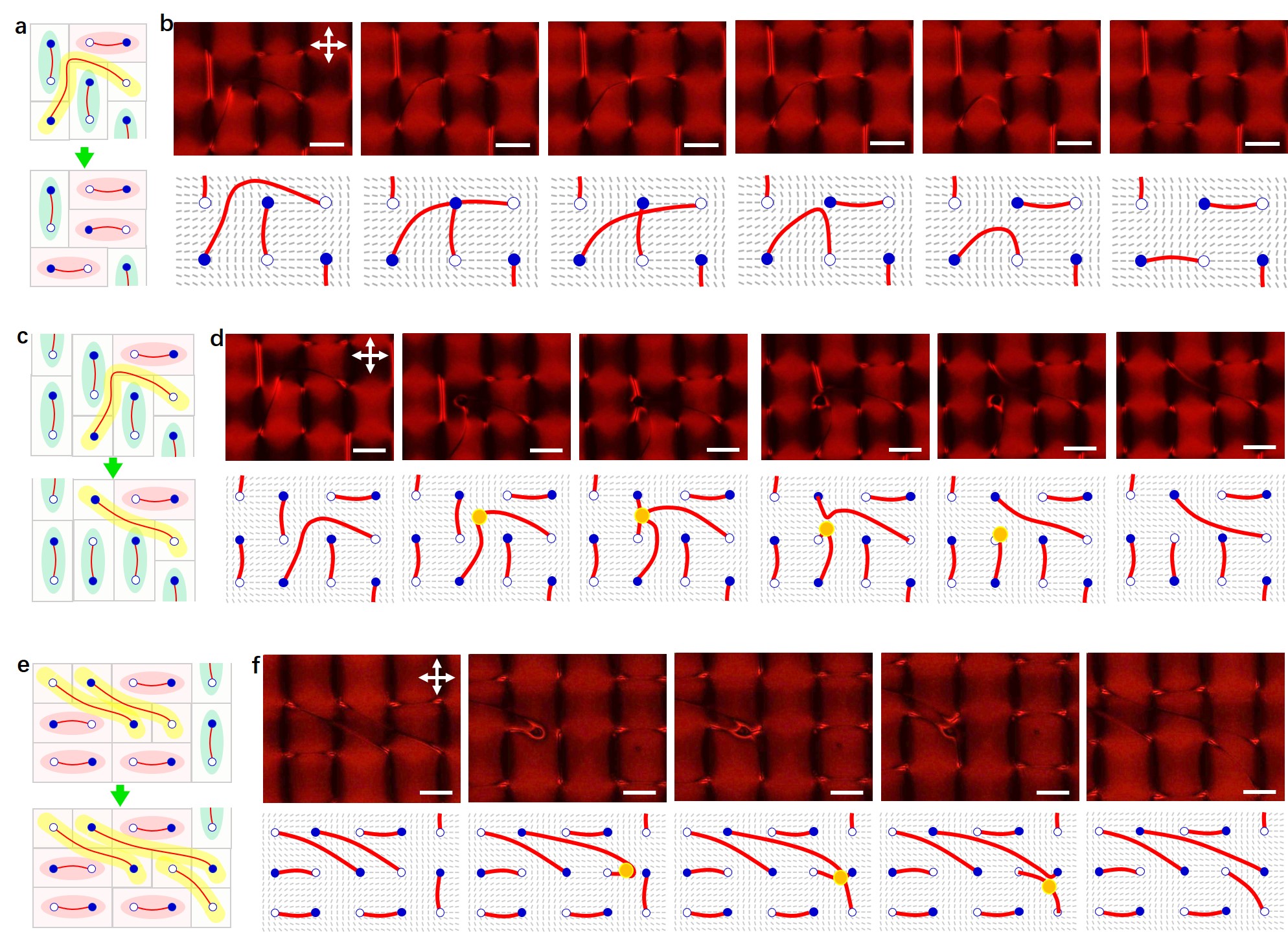}
\caption{\textbf{Topological surgery on dimers.} \textbf{a,b,} “Shrinkage” of dimer led by the minimization of LC elastic energy. The curved and elongated vortex line or dimer causes pronounced distortion in the nematic system, which causes the line to relax back to a straight one. During the relaxation, the line encounters another dimer and reconnects into a couple of lower-energy dimers. \textbf{c,d,} “Re-connection” of dimer in a 3 × 3 sub-lattice by rewiring of nearest neighbor dimers. \textbf{e,f,} Dimer “growing” as stretched over four pinning sites. 
 In (\textbf{a,c,e}), dimers arrangement before and after the topological surgery are schematically drawn. The green and pink ellipses denote dimers in vertical and horizontal directions, respectively. A yellow-highlighted line represents a diagonally wiring dimer. In (\textbf{b,d,f}), the upper row is experimental snapshots showing the dimer revolution without (\textbf{b}) or with (\textbf{d,f}) the need of laser tweezer. The lower row is schematic showing the corresponding dimer ensemble with LC director field \textbf{n(r)} beneath. Double white arrows mark the orientations of the crossed polarizers. All scale bars are \SI{50}{\micro\metre}.}
\end{figure*}

\begin{figure*}[th!!!!!]
  \includegraphics[width=\textwidth]{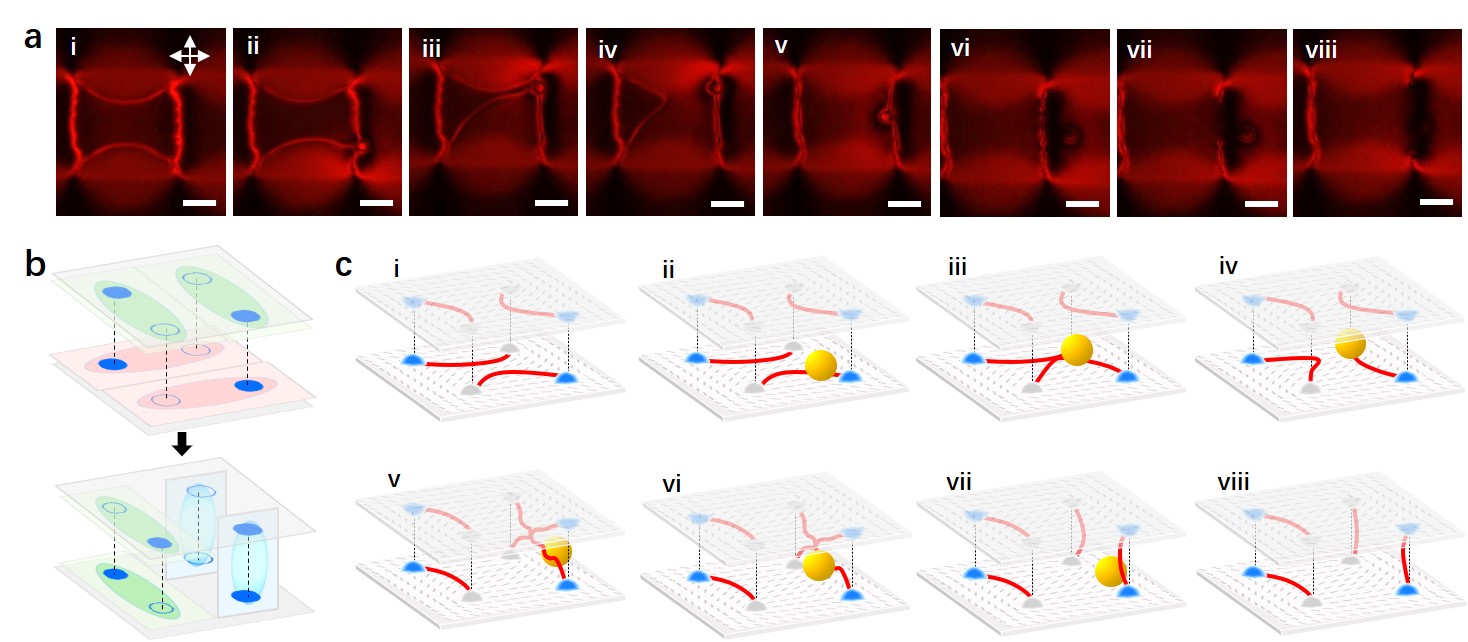}
\caption{\textbf{Three-dimensional LC dimer models.} \textbf{a, }A series of polarizing optical images showing how to realize a co-existing dimer ensemble having both longitudinal and horizontal dimers in a single cell. That is, two lines stand longitudinally by connecting the pinning sites at the top-bottom surfaces and two horizontal lines connect the pinning site in-plane at the top-top or bottom-bottom surfaces. \textbf{b}, Three-dimensional sketch showing the transition of dimer ensembles with pink, green and blue ellipses representing horizontal, vertical and longitudinal dimers, respectively. \textbf{c}, Schematic showing spatial revolution of the vortex lines. Double white arrows mark the orientations of the crossed polarizers. 
All scale bars are \SI{25}{\micro\metre}. 
   }
\end{figure*}

\begin{figure*}[th!]
  \includegraphics[width=\textwidth]{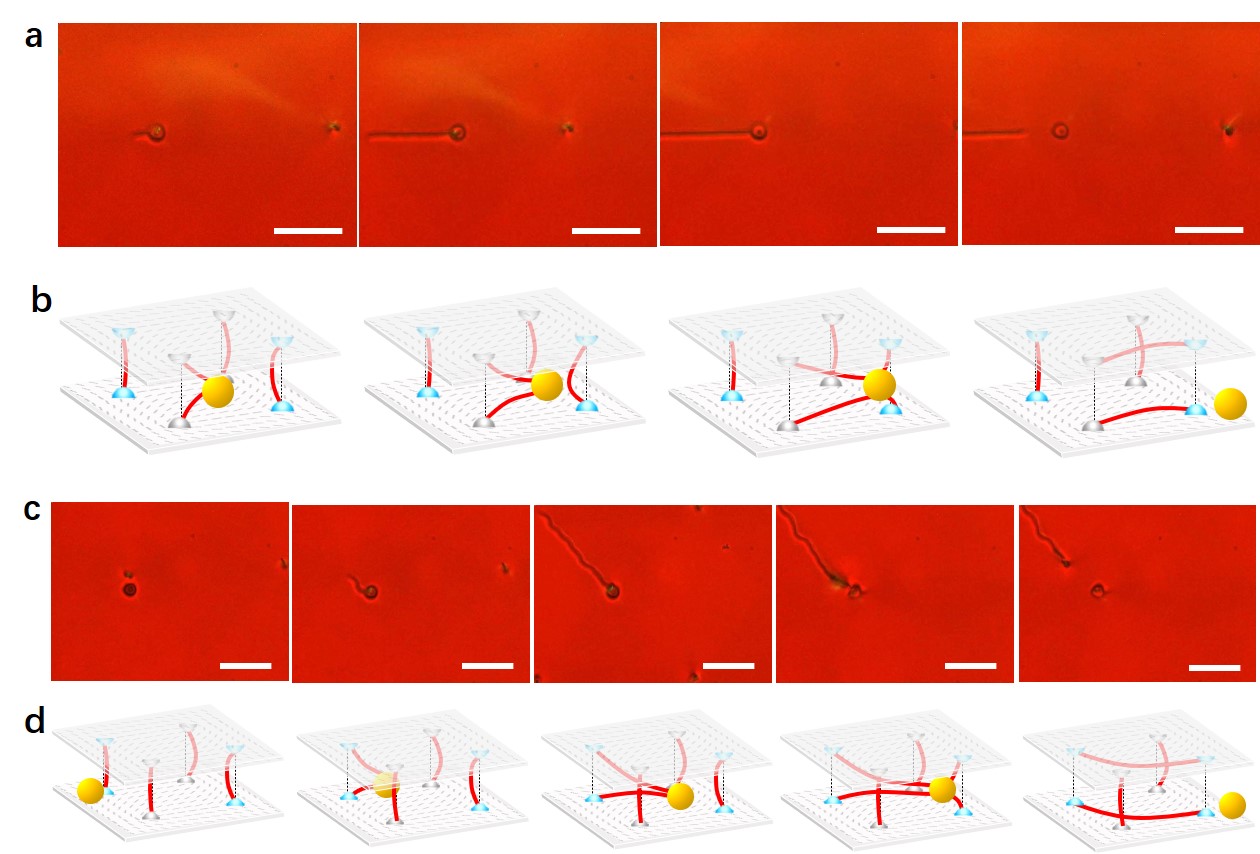}
\caption{\textbf{Optical manipulation of three-dimensional dimer models. } 
\textbf{a,c,} Snapshots of bright-field micrography showing intermediate trajectories of vortex lines from longitudinal to (\textbf{a}) in-plane horizontal or (\textbf{c}) diagonal connection using laser tweezer (shown as a dimmed dot), which demonstrates an inverse process of topological surgery shown in Fig. 3f. The background showing red colour is due to the insertion of a red filter to avoid unwanted effect from the imaging light. \textbf{b,d}\textbf{,} Schematic showing the corresponding steps. All scale bars are \SI{100}{\micro\metre}.   }
\end{figure*}

\begin{figure*}[th!]
  \includegraphics[width=\textwidth]{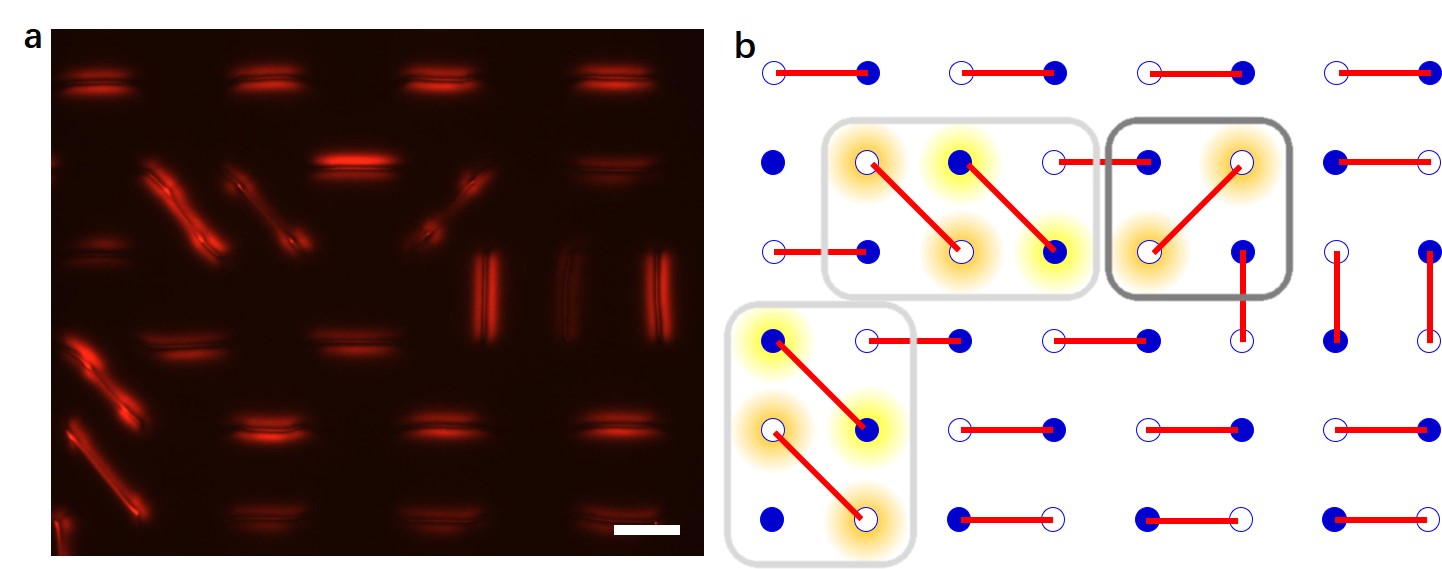}
\caption{\textbf{a,b,} Phase-contrast micrograph and schematic view of dimer ensemble with diagonal dimerization. The dimers in light-gray boxes in (\textbf{b}) can be eliminated with each other due to a zero net charge it encloses, while the dimers in the dark-grey box are not allowed to be removed since they contain a non-zero net charge. 
Scale bar is \SI{50}{\micro\metre}.}
\end{figure*}

We model the LC equilibrium structures by numerical minimization of the free energy~\cite{kos2022nematic}:
\begin{align*}
    F &= \int \mathrm{d}V \left( \frac{A}{2}Q_{ij}Q_{ji} + \frac{B}{3} Q_{ij}Q_{jk}Q_{ki} + \frac{C}{4} \left(Q_{ij}Q_{ji}\right)^2 \right.\\
    & + \left.\frac{1}{2} L (\partial_k Q_{ij})(\partial_k Q_{ij}) \right)
     + \frac{W}{2} \int \mathrm{d}S\left(Q_{ij}-Q_{ij}^0 \right)^2,
\end{align*}
where $Q$ is the tensorial order parameter, $A$, $B$, and $C$ are phase parameters, $L$ is the elastic constants, and $Q^0$ is the surface-preferred order parameter. 
The tensorial description of nematic order includes both the director as the main eigenvector of the Q-tensor, and the scalar degree of order as the main eigenvalue. Within defect lines, the scalar degree of order drops towards zero.
Equilibrium structures are found using gradient descent on a finite difference mesh. $Q^0$ profiles are prescribed at the top and the bottom surface and periodic boundary conditions are used at the side surfaces. 
The height of the numerical simulation box equals $20\Delta x$ in Figs.~\ref{fig:1} and \ref{fig:5} and $30\Delta x$ in Fig.~\ref{fig:4}, where $\Delta x$ is the mesh resolution.
The surface anchoring strength is varied between $W=0.1\,L/\Delta x$ and $W=3\,L/\Delta x$. Phase parameters are set to $A=-0.43\,L/(\Delta x)^2$, $B=-5.3\,L/(\Delta x)^2$, and $C=4.33\,L/(\Delta x)^2$. 
The numerical approach was used to simulate the nematic structure in Figs.~1b, 4b,d,f, 5d,j and also Fig.~13.

\vspace{10mm}

\begin{figure*}[tb!]
  \includegraphics[width=\textwidth]{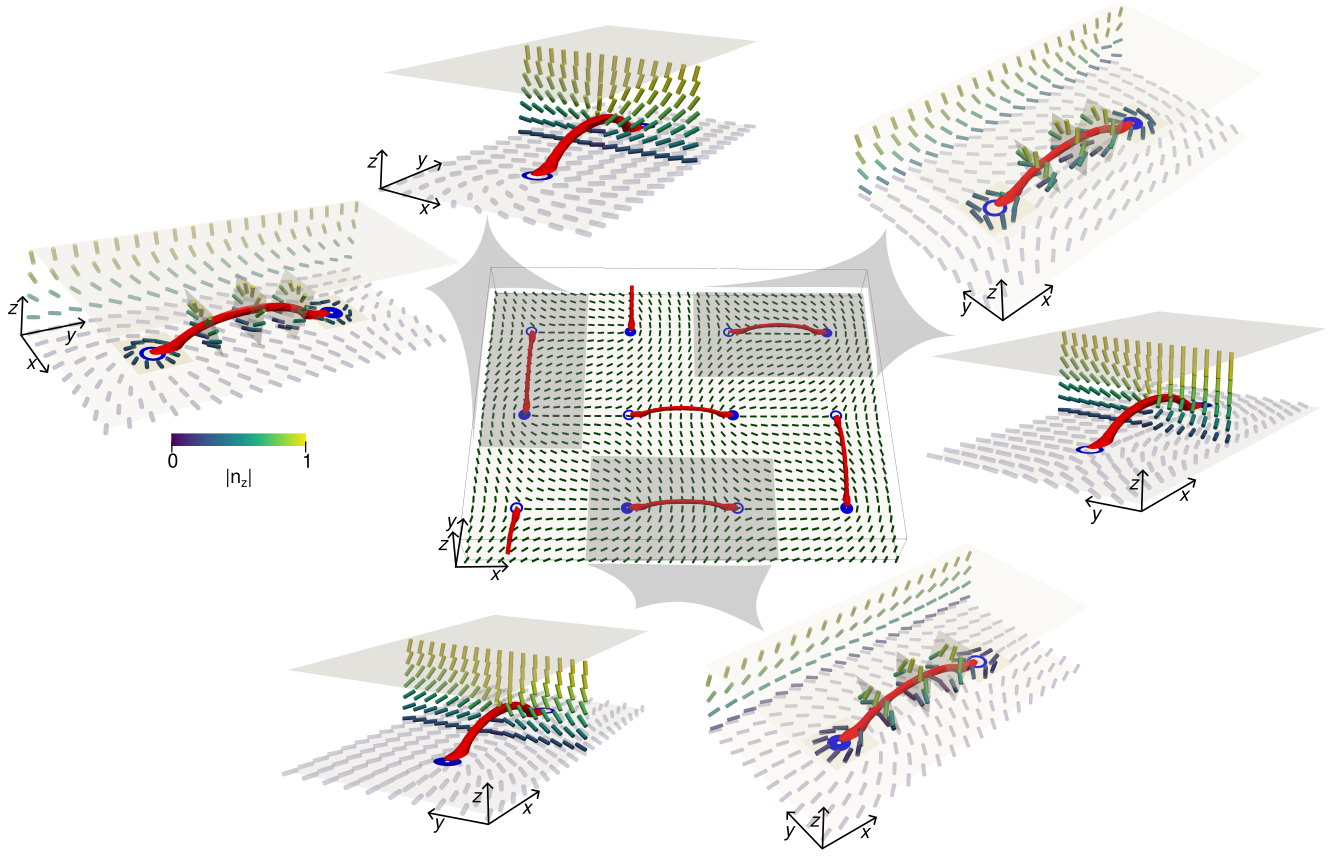}
  \caption{
  \textbf{Three-dimensional director structure of vortex lines in combinatorial lattices.} The director field is shown for one dimer oriented along $y$-direction and two dimers along $x$-direction. In all cases shown, the vortex line transitions from a patterned texture with a $+1/2$ singularity to a texture with a $-1/2$ singularity. In the middle of the dimer, the director field can have a wedge-like shape (bottom 2 panels), twisted shape (top right) or an intermediate shape (top left).} 
  
\end{figure*}

\textbf{Data availability}

All data are available in the main text or the supplementary materials.
The codebase used for generation of the director field for vortices lattices is available upon reasonable request.\vspace{5mm}
 
\textbf{Acknowledgements}

Research at CU-Boulder was supported by the U.S. Department of Energy, Office of Basic Energy Sciences, Division of Materials Sciences and Engineering, under contract DE-SC0019293 with the University of Colorado at Boulder (C.M., J.-S.W. and I.I.S.). I.I.S. also acknowledges the hospitality of the International Institute for Sustainability with Knotted Chiral Meta Matter (SKCM2) at the University of Hiroshima, Japan, during his sabbatical stay, where he was partly working on this article.  Ž.K. acknowledges funding from Slovenian Research and Innovation Agency (ARIS) under contracts P1-0099 and J1-50006. The work of C.N.  was carried out under the auspices of the U.S. DOE through LANL, operated by Triad National Security, LLC (Contract No. 892333218NCA000001) and financed by DOE, LDRD office. We thank T. Lee, B. Senyuk and H. Zhao for technical assistance and discussions. We thank Claudio Castelnovo (Cambridge), Stephen Powell (Notthingam), and Richard Kenyon (Yale) for feedback on the manuscript.\vspace{5mm}

\textbf{Author Contributions}

C.M. and J.-S.W. performed the experiments.  J.-S.W., C.M. and I.I.S. designed and experimentally created the lattices of nematic vortices and analysed the data. Ž.K. performed Q-tensor modeling. Ž.K. and C.N. designed and analyzed Dirac strings and other emergent defects in lattices of nematic vortices. C.N. and J.D. supervised the theoretical research and I.I.S. supervised the experimental research and design of lattices of vortices. C.M., Ž.K., C.N. and I.I.S. wrote the manuscript, with feedback and contributions from all authors. I.I.S. conceived the project and initiated the collaboration.

\clearpage


\begin{thebibliography}{10}
\expandafter\ifx\csname url\endcsname\relax
  \def\url#1{\texttt{#1}}\fi
\expandafter\ifx\csname urlprefix\endcsname\relax\def\urlprefix{URL }\fi
\providecommand{\bibinfo}[2]{#2}
\providecommand{\eprint}[2][]{\url{#2}}

\bibitem{topophys}
\bibinfo{author}{Yang, C.~N.}, \bibinfo{author}{Ge, M.-L.} \&
  \bibinfo{author}{He, Y.-H.}
\newblock \emph{\bibinfo{title}{Topology and Physics}}
  (\bibinfo{publisher}{World Scientific}, \bibinfo{year}{2019}).

\bibitem{knotphys}
\bibinfo{author}{Kauffman, L.~H.}
\newblock \emph{\bibinfo{title}{Knots and Physics}} (\bibinfo{publisher}{World
  Scientific}, \bibinfo{year}{2001}).

\bibitem{smalyukh2020knots}
\bibinfo{author}{Smalyukh, I.~I.}
\newblock \bibinfo{title}{Knots and other new topological effects in liquid
  crystals and colloids}.
\newblock \emph{\bibinfo{journal}{Reports on Progress in Physics}}
  \textbf{\bibinfo{volume}{83}}, \bibinfo{pages}{106601}
  (\bibinfo{year}{2020}).

\bibitem{softmatterphys}
\bibinfo{author}{Kleman, M.} \& \bibinfo{author}{Lavrentovich, O.~D.}
\newblock \emph{\bibinfo{title}{Soft Matter Physics:An Introduction}}
  (\bibinfo{publisher}{Springer}, \bibinfo{year}{2001}).

\bibitem{anderson1987resonating}
\bibinfo{author}{Anderson, P.~W.}, \bibinfo{author}{Baskaran, G.},
  \bibinfo{author}{Zou, Z.} \& \bibinfo{author}{Hsu, T.}
\newblock \bibinfo{title}{Resonating--valence-bond theory of phase transitions
  and superconductivity in la 2 cuo 4-based compounds}.
\newblock \emph{\bibinfo{journal}{Physical review letters}}
  \textbf{\bibinfo{volume}{58}}, \bibinfo{pages}{2790} (\bibinfo{year}{1987}).

\bibitem{moessner2010quantum}
\bibinfo{author}{Moessner, R.} \& \bibinfo{author}{Raman, K.~S.}
\newblock \bibinfo{title}{Quantum dimer models}.
\newblock In \emph{\bibinfo{booktitle}{Introduction to frustrated magnetism:
  materials, experiments, theory}}, \bibinfo{pages}{437--479}
  (\bibinfo{publisher}{Springer}, \bibinfo{year}{2010}).

\bibitem{laughlin1983anomalous}
\bibinfo{author}{Laughlin, R.~B.}
\newblock \bibinfo{title}{Anomalous quantum hall effect: an incompressible
  quantum fluid with fractionally charged excitations}.
\newblock \emph{\bibinfo{journal}{Physical Review Letters}}
  \textbf{\bibinfo{volume}{50}}, \bibinfo{pages}{1395} (\bibinfo{year}{1983}).

\bibitem{wen1989vacuum}
\bibinfo{author}{Wen, X.-G.}
\newblock \bibinfo{title}{Vacuum degeneracy of chiral spin states in
  compactified space}.
\newblock \emph{\bibinfo{journal}{Physical Review B}}
  \textbf{\bibinfo{volume}{40}}, \bibinfo{pages}{7387} (\bibinfo{year}{1989}).

\bibitem{sachdev2018topological}
\bibinfo{author}{Sachdev, S.}
\newblock \bibinfo{title}{Topological order, emergent gauge fields, and fermi
  surface reconstruction}.
\newblock \emph{\bibinfo{journal}{Reports on Progress in Physics}}
  \textbf{\bibinfo{volume}{82}}, \bibinfo{pages}{014001}
  (\bibinfo{year}{2018}).

\bibitem{wen2002quantum}
\bibinfo{author}{Wen, X.-G.}
\newblock \bibinfo{title}{Quantum orders and symmetric spin liquids}.
\newblock \emph{\bibinfo{journal}{Physical Review B}}
  \textbf{\bibinfo{volume}{65}}, \bibinfo{pages}{165113}
  (\bibinfo{year}{2002}).

\bibitem{nayak2008non}
\bibinfo{author}{Nayak, C.}, \bibinfo{author}{Simon, S.~H.},
  \bibinfo{author}{Stern, A.}, \bibinfo{author}{Freedman, M.} \&
  \bibinfo{author}{Sarma, S.~D.}
\newblock \bibinfo{title}{Non-abelian anyons and topological quantum
  computation}.
\newblock \emph{\bibinfo{journal}{Reviews of Modern Physics}}
  \textbf{\bibinfo{volume}{80}}, \bibinfo{pages}{1083} (\bibinfo{year}{2008}).

\bibitem{henley2010coulomb}
\bibinfo{author}{Henley, C.~L.}
\newblock \bibinfo{title}{The coulomb phase in frustrated systems}.
\newblock \emph{\bibinfo{journal}{Annu. Rev. Condens. Matter Phys.}}
  \textbf{\bibinfo{volume}{1}}, \bibinfo{pages}{179--210}
  (\bibinfo{year}{2010}).

\bibitem{henley2011classical}
\bibinfo{author}{Henley, C.~L.}
\newblock \bibinfo{title}{Classical height models with topological order}.
\newblock \emph{\bibinfo{journal}{Journal of Physics: Condensed Matter}}
  \textbf{\bibinfo{volume}{23}}, \bibinfo{pages}{164212}
  (\bibinfo{year}{2011}).

\bibitem{macdonald2011classical}
\bibinfo{author}{Macdonald, A.~J.}, \bibinfo{author}{Holdsworth, P.~C.} \&
  \bibinfo{author}{Melko, R.~G.}
\newblock \bibinfo{title}{Classical topological order in kagome ice}.
\newblock \emph{\bibinfo{journal}{Journal of Physics: Condensed Matter}}
  \textbf{\bibinfo{volume}{23}}, \bibinfo{pages}{164208}
  (\bibinfo{year}{2011}).

\bibitem{castelnovo2012spin}
\bibinfo{author}{Castelnovo, C.}, \bibinfo{author}{Moessner, R.} \&
  \bibinfo{author}{Sondhi, S.}
\newblock \bibinfo{title}{Spin ice, fractionalization, and topological order}.
\newblock \emph{\bibinfo{journal}{Annu. Rev. Condens. Matter Phys.}}
  \textbf{\bibinfo{volume}{3}}, \bibinfo{pages}{35--55} (\bibinfo{year}{2012}).

\bibitem{jaubert2013topological}
\bibinfo{author}{Jaubert, L.~D.} \emph{et~al.}
\newblock \bibinfo{title}{Topological-sector fluctuations and curie-law
  crossover in spin ice}.
\newblock \emph{\bibinfo{journal}{Physical Review X}}
  \textbf{\bibinfo{volume}{3}}, \bibinfo{pages}{011014} (\bibinfo{year}{2013}).

\bibitem{nisoli2020topological}
\bibinfo{author}{Nisoli, C.}
\newblock \bibinfo{title}{Topological order of the rys f-model and its
  breakdown in realistic square spin ice: Topological sectors of faraday
  loops}.
\newblock \emph{\bibinfo{journal}{EPL (Europhysics Letters)}}
  \textbf{\bibinfo{volume}{132}}, \bibinfo{pages}{47005}
  (\bibinfo{year}{2020}).

\bibitem{zhang2023topological}
\bibinfo{author}{Zhang, X.} \emph{et~al.}
\newblock \bibinfo{title}{Topological kinetic crossover in a nanomagnet array}.
\newblock \emph{\bibinfo{journal}{Science}} \textbf{\bibinfo{volume}{380}},
  \bibinfo{pages}{526--531} (\bibinfo{year}{2023}).

\bibitem{kivelson1987topology}
\bibinfo{author}{Kivelson, S.~A.}, \bibinfo{author}{Rokhsar, D.~S.} \&
  \bibinfo{author}{Sethna, J.~P.}
\newblock \bibinfo{title}{Topology of the resonating valence-bond state:
  Solitons and high-t c superconductivity}.
\newblock \emph{\bibinfo{journal}{Physical Review B}}
  \textbf{\bibinfo{volume}{35}}, \bibinfo{pages}{8865} (\bibinfo{year}{1987}).

\bibitem{moessner2001resonating}
\bibinfo{author}{Moessner, R.} \& \bibinfo{author}{Sondhi, S.~L.}
\newblock \bibinfo{title}{Resonating valence bond phase in the triangular
  lattice quantum dimer model}.
\newblock \emph{\bibinfo{journal}{Physical Review Letters}}
  \textbf{\bibinfo{volume}{86}}, \bibinfo{pages}{1881} (\bibinfo{year}{2001}).

\bibitem{di2022memcomputing}
\bibinfo{author}{Di~Ventra, M.}
\newblock \emph{\bibinfo{title}{MemComputing: fundamentals and applications}}
  (\bibinfo{publisher}{Oxford University Press}, \bibinfo{year}{2022}).

\bibitem{gartside2022reconfigurable}
\bibinfo{author}{Gartside, J.~C.} \emph{et~al.}
\newblock \bibinfo{title}{Reconfigurable training and reservoir computing in an
  artificial spin-vortex ice via spin-wave fingerprinting}.
\newblock \emph{\bibinfo{journal}{Nature Nanotechnology}}
  \textbf{\bibinfo{volume}{17}}, \bibinfo{pages}{460--469}
  (\bibinfo{year}{2022}).

\bibitem{PetitGarrido}
\bibinfo{author}{Petit-Garrido, N.} \emph{et~al.}
\newblock \bibinfo{title}{Healing of defects at the interface of nematic liquid
  crystals and structured langmuir-blodgett monolayers}.
\newblock \emph{\bibinfo{journal}{Physical Review Letters}}
  \textbf{\bibinfo{volume}{107}}, \bibinfo{pages}{177801}
  (\bibinfo{year}{2011}).

\bibitem{perrin2016extensive}
\bibinfo{author}{Perrin, Y.}, \bibinfo{author}{Canals, B.} \&
  \bibinfo{author}{Rougemaille, N.}
\newblock \bibinfo{title}{Extensive degeneracy, coulomb phase and magnetic
  monopoles in artificial square ice}.
\newblock \emph{\bibinfo{journal}{Nature}} \textbf{\bibinfo{volume}{540}},
  \bibinfo{pages}{410--413} (\bibinfo{year}{2016}).

\bibitem{lao2018classical}
\bibinfo{author}{Lao, Y.} \emph{et~al.}
\newblock \bibinfo{title}{Classical topological order in the kinetics of
  artificial spin ice}.
\newblock \emph{\bibinfo{journal}{Nature Physics}}
  \textbf{\bibinfo{volume}{14}}, \bibinfo{pages}{723--727}
  (\bibinfo{year}{2018}).

\bibitem{libal2018ice}
\bibinfo{author}{Lib{\'a}l, A.} \emph{et~al.}
\newblock \bibinfo{title}{Ice rule fragility via topological charge transfer in
  artificial colloidal ice}.
\newblock \emph{\bibinfo{journal}{Nature communications}}
  \textbf{\bibinfo{volume}{9}}, \bibinfo{pages}{4146} (\bibinfo{year}{2018}).

\bibitem{chen2019experimental}
\bibinfo{author}{Chen, T.} \emph{et~al.}
\newblock \bibinfo{title}{Experimental observation of classical analogy of
  topological entanglement entropy}.
\newblock \emph{\bibinfo{journal}{Nature communications}}
  \textbf{\bibinfo{volume}{10}}, \bibinfo{pages}{1557} (\bibinfo{year}{2019}).

\bibitem{sirote2024emergent}
\bibinfo{author}{Sirote-Katz, C.} \emph{et~al.}
\newblock \bibinfo{title}{Emergent disorder and mechanical memory in periodic
  metamaterials}.
\newblock \emph{\bibinfo{journal}{Nature Communications}}
  \textbf{\bibinfo{volume}{15}}, \bibinfo{pages}{4008} (\bibinfo{year}{2024}).

\bibitem{king2021qubit}
\bibinfo{author}{King, A.~D.}, \bibinfo{author}{Nisoli, C.},
  \bibinfo{author}{Dahl, E.~D.}, \bibinfo{author}{Poulin-Lamarre, G.} \&
  \bibinfo{author}{Lopez-Bezanilla, A.}
\newblock \bibinfo{title}{Qubit spin ice}.
\newblock \emph{\bibinfo{journal}{Science}}  (\bibinfo{year}{2021}).

\bibitem{lopez2023kagome}
\bibinfo{author}{Lopez-Bezanilla, A.} \emph{et~al.}
\newblock \bibinfo{title}{Kagome qubit ice}.
\newblock \emph{\bibinfo{journal}{Nature Communications}}
  \textbf{\bibinfo{volume}{14}}, \bibinfo{pages}{1105} (\bibinfo{year}{2023}).

\bibitem{duzgun2021skyrmion}
\bibinfo{author}{Duzgun, A.} \& \bibinfo{author}{Nisoli, C.}
\newblock \bibinfo{title}{Skyrmion spin ice in liquid crystals}.
\newblock \emph{\bibinfo{journal}{Physical Review Letters}}
  \textbf{\bibinfo{volume}{126}}, \bibinfo{pages}{047801}
  (\bibinfo{year}{2021}).

\bibitem{tai2023field}
\bibinfo{author}{Tai, J.-S.~B.}, \bibinfo{author}{Hess, A.~J.},
  \bibinfo{author}{Wu, J.-S.} \& \bibinfo{author}{Smalyukh, I.~I.}
\newblock \bibinfo{title}{Field-controlled dynamics of skyrmions and
  monopoles}.
\newblock \emph{\bibinfo{journal}{Science Advances}}
  \textbf{\bibinfo{volume}{10}}, \bibinfo{pages}{adj9373}
  (\bibinfo{year}{2024}).

\bibitem{kasteleyn1961statistics}
\bibinfo{author}{Kasteleyn, P.~W.}
\newblock \bibinfo{title}{The statistics of dimers on a lattice: I. the number
  of dimer arrangements on a quadratic lattice}.
\newblock \emph{\bibinfo{journal}{Physica}} \textbf{\bibinfo{volume}{27}},
  \bibinfo{pages}{1209--1225} (\bibinfo{year}{1961}).

\bibitem{Baxter1982}
\bibinfo{author}{Baxter, R.}
\newblock \emph{\bibinfo{title}{{Exactly solved models in statistical
  mechanics}}} (\bibinfo{publisher}{Academic}, \bibinfo{address}{New York},
  \bibinfo{year}{1982}).

\bibitem{Martinez}
\bibinfo{author}{Martinez, A.}, \bibinfo{author}{Mireles, H.~C.} \&
  \bibinfo{author}{Smalyukh, I.~I.}
\newblock \bibinfo{title}{Large-area optoelastic manipulation of colloidal
  particles in liquid crystals using photoresponsive molecular surface
  monolayers}.
\newblock \emph{\bibinfo{journal}{PROC. NATL. ACAD. SCI. U.S.A.}}
  \textbf{\bibinfo{volume}{108}}, \bibinfo{pages}{20891}
  (\bibinfo{year}{2011}).

\bibitem{meng2023topological}
\bibinfo{author}{Meng, C.}, \bibinfo{author}{Wu, J.-S.} \&
  \bibinfo{author}{Smalyukh, I.~I.}
\newblock \bibinfo{title}{Topological steering of light by nematic vortices and
  analogy to cosmic strings}.
\newblock \emph{\bibinfo{journal}{Nature Materials}}
  \textbf{\bibinfo{volume}{22}}, \bibinfo{pages}{64--72}
  (\bibinfo{year}{2023}).

\bibitem{kenyonintroduction}
\bibinfo{author}{Kenyon, R.}
\newblock \bibinfo{title}{An introduction to the dimer model}.
\newblock In \emph{\bibinfo{booktitle}{School and Conference on Probability
  Theory}}, vol.~\bibinfo{volume}{17}, \bibinfo{pages}{267}
  (\bibinfo{publisher}{ICTP Lecture Notes Series}, \bibinfo{year}{2004}).

\bibitem{meng2018hybridcell}
\bibinfo{author}{Meng, C.}, \bibinfo{author}{Tseng, M.}, \bibinfo{author}{Tang,
  S.} \& \bibinfo{author}{Kwok, H.-S.}
\newblock \bibinfo{title}{Optical rewritable liquid crystal displays without a
  front polarizer}.
\newblock \emph{\bibinfo{journal}{Optics Letters}}
  \textbf{\bibinfo{volume}{43}}, \bibinfo{pages}{899--902}
  (\bibinfo{year}{2018}).

\bibitem{oakes2016emergence}
\bibinfo{author}{Oakes, T.}, \bibinfo{author}{Garrahan, J.~P.} \&
  \bibinfo{author}{Powell, S.}
\newblock \bibinfo{title}{Emergence of cooperative dynamics in fully packed
  classical dimers}.
\newblock \emph{\bibinfo{journal}{Physical Review E}}
  \textbf{\bibinfo{volume}{93}}, \bibinfo{pages}{032129}
  (\bibinfo{year}{2016}).

\bibitem{henley1997relaxation}
\bibinfo{author}{Henley, C.~L.}
\newblock \bibinfo{title}{Relaxation time for a dimer covering with height
  representation}.
\newblock \emph{\bibinfo{journal}{Journal of statistical physics}}
  \textbf{\bibinfo{volume}{89}}, \bibinfo{pages}{483--507}
  (\bibinfo{year}{1997}).

\bibitem{antoniou2017extending}
\bibinfo{author}{Antoniou, S.} \& \bibinfo{author}{Lambropoulou, S.}
\newblock \bibinfo{title}{Extending topological surgery to natural processes
  and dynamical systems}.
\newblock \emph{\bibinfo{journal}{PLoS one}} \textbf{\bibinfo{volume}{12}},
  \bibinfo{pages}{e0183993} (\bibinfo{year}{2017}).

\bibitem{Castelnovo2008}
\bibinfo{author}{Castelnovo, C.}, \bibinfo{author}{Moessner, R.} \&
  \bibinfo{author}{Sondhi, S.~L.}
\newblock \bibinfo{title}{{Magnetic monopoles in spin ice}}.
\newblock \emph{\bibinfo{journal}{Nature}} \textbf{\bibinfo{volume}{451}},
  \bibinfo{pages}{42--5} (\bibinfo{year}{2008}).

\bibitem{desai2021bilayer}
\bibinfo{author}{Desai, N.}, \bibinfo{author}{Pujari, S.} \&
  \bibinfo{author}{Damle, K.}
\newblock \bibinfo{title}{Bilayer coulomb phase of two-dimensional dimer
  models: Absence of power-law columnar order}.
\newblock \emph{\bibinfo{journal}{Physical Review E}}
  \textbf{\bibinfo{volume}{103}}, \bibinfo{pages}{042136}
  (\bibinfo{year}{2021}).

\bibitem{kenyon2014conformal}
\bibinfo{author}{Kenyon, R.}
\newblock \bibinfo{title}{Conformal invariance of loops in the double-dimer
  model}.
\newblock \emph{\bibinfo{journal}{Communications in Mathematical Physics}}
  \textbf{\bibinfo{volume}{326}}, \bibinfo{pages}{477--497}
  (\bibinfo{year}{2014}).

\bibitem{kenyon2010double}
\bibinfo{author}{Kenyon, R.~W.} \& \bibinfo{author}{Wilson, D.~B.}
\newblock \bibinfo{title}{Double-dimer pairings and skew young diagrams}.
\newblock \emph{\bibinfo{journal}{arXiv preprint arXiv:1007.2006}}
  (\bibinfo{year}{2010}).

\bibitem{kenyon2011boundary}
\bibinfo{author}{Kenyon, R.~W.} \& \bibinfo{author}{Wilson, D.}
\newblock \bibinfo{title}{Boundary partitions in trees and dimers}.
\newblock \emph{\bibinfo{journal}{Transactions of the American Mathematical
  Society}} \textbf{\bibinfo{volume}{363}}, \bibinfo{pages}{1325--1364}
  (\bibinfo{year}{2011}).

\bibitem{lieb1972inphase}
\bibinfo{author}{Lieb, E.} \& \bibinfo{author}{Wu, F.}
\newblock \bibinfo{title}{Two-dimensional ferroelectric models}
  (\bibinfo{year}{1972}).

\bibitem{caravelli2020logical}
\bibinfo{author}{Caravelli, F.} \& \bibinfo{author}{Nisoli, C.}
\newblock \bibinfo{title}{Logical gates embedding in artificial spin ice}.
\newblock \emph{\bibinfo{journal}{New Journal of Physics}}
  \textbf{\bibinfo{volume}{22}}, \bibinfo{pages}{103052}
  (\bibinfo{year}{2020}).

\bibitem{pivsljar2022blue}
\bibinfo{author}{Pi{\v{s}}ljar, J.} \emph{et~al.}
\newblock \bibinfo{title}{Blue phase iii: topological fluid of skyrmions}.
\newblock \emph{\bibinfo{journal}{Physical Review X}}
  \textbf{\bibinfo{volume}{12}}, \bibinfo{pages}{011003}
  (\bibinfo{year}{2022}).

\bibitem{ackerman2017diversity}
\bibinfo{author}{Ackerman, P.~J.} \& \bibinfo{author}{Smalyukh, I.~I.}
\newblock \bibinfo{title}{Diversity of knot solitons in liquid crystals
  manifested by linking of preimages in torons and hopfions}.
\newblock \emph{\bibinfo{journal}{Physical Review X}}
  \textbf{\bibinfo{volume}{7}}, \bibinfo{pages}{011006} (\bibinfo{year}{2017}).

\bibitem{kim1994critical}
\bibinfo{author}{Kim, M.~H.} \& \bibinfo{author}{Park, H.}
\newblock \bibinfo{title}{Critical behavior of an interacting monomer-dimer
  model}.
\newblock \emph{\bibinfo{journal}{Physical review letters}}
  \textbf{\bibinfo{volume}{73}}, \bibinfo{pages}{2579} (\bibinfo{year}{1994}).

\bibitem{alet2005interacting}
\bibinfo{author}{Alet, F.} \emph{et~al.}
\newblock \bibinfo{title}{Interacting classical dimers on the square lattice}.
\newblock \emph{\bibinfo{journal}{Physical review letters}}
  \textbf{\bibinfo{volume}{94}}, \bibinfo{pages}{235702}
  (\bibinfo{year}{2005}).

\bibitem{wilkins2020interacting}
\bibinfo{author}{Wilkins, N.} \& \bibinfo{author}{Powell, S.}
\newblock \bibinfo{title}{Interacting double dimer model on the square
  lattice}.
\newblock \emph{\bibinfo{journal}{Physical Review B}}
  \textbf{\bibinfo{volume}{102}}, \bibinfo{pages}{174431}
  (\bibinfo{year}{2020}).

\bibitem{barahona1982computational}
\bibinfo{author}{Barahona, F.}
\newblock \bibinfo{title}{On the computational complexity of ising spin glass
  models}.
\newblock \emph{\bibinfo{journal}{Journal of Physics A: Mathematical and
  General}} \textbf{\bibinfo{volume}{15}}, \bibinfo{pages}{3241}
  (\bibinfo{year}{1982}).

\bibitem{FerroSwitching}
\bibinfo{author}{Zhang, Q.}, \bibinfo{author}{Ackerman, P.},
  \bibinfo{author}{Liu, Q.} \& \bibinfo{author}{Smalyukh, I.}
\newblock \bibinfo{title}{Ferromagnetic switching of knotted vector fields in
  liquid crystal colloids}.
\newblock \emph{\bibinfo{journal}{Physical Review Letters}}
  \textbf{\bibinfo{volume}{115}}, \bibinfo{pages}{097802}
  (\bibinfo{year}{2015}).

\bibitem{wei2014optvortices}
\bibinfo{author}{Wei, B.} \emph{et~al.}
\newblock \bibinfo{title}{Generating switchable and reconﬁgurable optical
  vortices via photopatterning of liquid crystals}.
\newblock \emph{\bibinfo{journal}{Advanced Materials}}
  \textbf{\bibinfo{volume}{26}}, \bibinfo{pages}{1590--1595}
  (\bibinfo{year}{2014}).

\bibitem{nys2022photoaligned}
\bibinfo{author}{Nys, I.}, \bibinfo{author}{Berteloot, B.} \&
  \bibinfo{author}{Neyts, K.}
\newblock \bibinfo{title}{Photoaligned chiral liquid crystal grating with
  hysteresis switching}.
\newblock \emph{\bibinfo{journal}{Advanced Optical Materials}}
  \textbf{\bibinfo{volume}{10}}, \bibinfo{pages}{2201289}
  (\bibinfo{year}{2022}).

\bibitem{jiang2024moire}
\bibinfo{author}{Wang, X.} \emph{et~al.}
\newblock \bibinfo{title}{Moir{\'e} effect enables versatile design of
  topological defects in nematic liquid crystals}.
\newblock \emph{\bibinfo{journal}{Nature Commuications}}
  \textbf{\bibinfo{volume}{15}}, \bibinfo{pages}{1655} (\bibinfo{year}{2024}).

\bibitem{mcginty2021large}
\bibinfo{author}{Mcginty, C.~P.}, \bibinfo{author}{Kolacz, J.} \&
  \bibinfo{author}{Spillmann, C.~M.}
\newblock \bibinfo{title}{Large rewritable liquid crystal pretilt angle by in
  situ photoalignment of brilliant yellow films}.
\newblock \emph{\bibinfo{journal}{Applied physics letters}}
  \textbf{\bibinfo{volume}{119}}, \bibinfo{pages}{141111}
  (\bibinfo{year}{2021}).

\bibitem{Neyts2018PAtoVA}
\bibinfo{author}{Nys, I.}, \bibinfo{author}{Chen, K.},
  \bibinfo{author}{Beeckman, J.} \& \bibinfo{author}{Neyts, K.}
\newblock \bibinfo{title}{Periodic planar-homeotropic anchoring realized by
  photoalignment for stabilization of chiral superstructures}.
\newblock \emph{\bibinfo{journal}{Advanced Optical Materials}}
  \textbf{\bibinfo{volume}{6}}, \bibinfo{pages}{1701163}
  (\bibinfo{year}{2018}).

\bibitem{gorkunov2020LCmetasurfaces}
\bibinfo{author}{Gorkunov, M.~V.} \emph{et~al.}
\newblock \bibinfo{title}{Liquid-crystal metasurfaces self-assembled on focused
  ion beam patterned polymer layers: Electro-optical control of light
  diﬀraction and transmission}.
\newblock \emph{\bibinfo{journal}{ACS applied materials \& interfaces}}
  \textbf{\bibinfo{volume}{12}}, \bibinfo{pages}{30815--30823}
  (\bibinfo{year}{2020}).

\bibitem{neyts2024multiimpose}
\bibinfo{author}{Berteloot, B.}, \bibinfo{author}{Nys, I.},
  \bibinfo{author}{Liu, S.} \& \bibinfo{author}{Neyts, K.}
\newblock \bibinfo{title}{Two-dimensional liquid-crystal photoalignment by
  multiple illumination steps}.
\newblock \emph{\bibinfo{journal}{ACS Applied Optical Materials}}
  \textbf{\bibinfo{volume}{2}}, \bibinfo{pages}{1295--1302}
  (\bibinfo{year}{2024}).

\bibitem{Zhi2023lightpulse}
\bibinfo{author}{Meng, Z.} \& \bibinfo{author}{Huang, W.}
\newblock \bibinfo{title}{Large aperture and defect-free liquid crystal planar
  optics enabled by high-throughput pulsed polarization patterning}.
\newblock \emph{\bibinfo{journal}{Optics Express}}
  \textbf{\bibinfo{volume}{31}}, \bibinfo{pages}{30435--30445}
  (\bibinfo{year}{2023}).

\bibitem{Egor2013nano}
\bibinfo{author}{Shteyner, E.}, \bibinfo{author}{Srivastava, A.},
  \bibinfo{author}{Chigrinov, V.}, \bibinfo{author}{Kwok, H.-S.} \&
  \bibinfo{author}{Afanasyev, A.}
\newblock \bibinfo{title}{Submicron-scale liquid crystal photo-alignment}.
\newblock \emph{\bibinfo{journal}{Soft Matter}} \textbf{\bibinfo{volume}{9}},
  \bibinfo{pages}{5160--5165} (\bibinfo{year}{2013}).

\bibitem{wu2024natcom}
\bibinfo{author}{Wu, J.-S.}, \bibinfo{author}{Torres~Lázaro, M.},
  \bibinfo{author}{Mundoor, H.}, \bibinfo{author}{Wensink, H.~H.} \&
  \bibinfo{author}{Smalyukh, I.~I.}
\newblock \bibinfo{title}{Emergent biaxiality in chiral hybrid liquid
  crystals}.
\newblock \emph{\bibinfo{journal}{Nature Commuications}}
  \textbf{\bibinfo{volume}{15}}, \bibinfo{pages}{9941} (\bibinfo{year}{2024}).

\bibitem{mundoor2021nat}
\bibinfo{author}{Mundoor, H.}, \bibinfo{author}{Wu, J.-S.},
  \bibinfo{author}{Wensink, H.~H.} \& \bibinfo{author}{Smalyukh, I.~I.}
\newblock \bibinfo{title}{Thermally reconfigurable monoclinic nematic colloidal
  fluids}.
\newblock \emph{\bibinfo{journal}{Nature}} \textbf{\bibinfo{volume}{590}},
  \bibinfo{pages}{268--274} (\bibinfo{year}{2021}).

\bibitem{mundoor2018sci}
\bibinfo{author}{Mundoor, H.}, \bibinfo{author}{Park, S.},
  \bibinfo{author}{Senyuk, B.}, \bibinfo{author}{Wensink, H.~H.} \&
  \bibinfo{author}{Smalyukh, I.~I.}
\newblock \bibinfo{title}{Hybrid molecular-colloidal liquid crystals}.
\newblock \emph{\bibinfo{journal}{Science}} \textbf{\bibinfo{volume}{360}},
  \bibinfo{pages}{768--771} (\bibinfo{year}{2018}).

\bibitem{liu2016pnas}
\bibinfo{author}{Liu, Q.}, \bibinfo{author}{Ackerman, P.~J.},
  \bibinfo{author}{Lubensky, T.~C.} \& \bibinfo{author}{Smalyukh, I.~I.}
\newblock \bibinfo{title}{Biaxial ferromagnetic liquid crystal colloids}.
\newblock \emph{\bibinfo{journal}{Proc. Natl. Acad. Sci. U.S.A.}}
  \textbf{\bibinfo{volume}{3}}, \bibinfo{pages}{10479--10484}
  (\bibinfo{year}{2016}).

\bibitem{wu2024arxiv}
\bibinfo{author}{Wu, J.-S.}, \bibinfo{author}{Valenzuela, R.~A.},
  \bibinfo{author}{Bowick, M.~J.} \& \bibinfo{author}{Smalyukh, I.~I.}
\newblock \bibinfo{title}{Topological rigidity and non-abelian defect junctions
  in chiral nematic systems with effective biaxial symmetry}.
\newblock \emph{\bibinfo{journal}{arXiv:2410.19293}}  (\bibinfo{year}{2024}).

\bibitem{tsung2021crystal}
\bibinfo{author}{Tsung, J.-W.}, \bibinfo{author}{Wang, Y.-Z.},
  \bibinfo{author}{Yao, S.-K.} \& \bibinfo{author}{Chao, S.-Y.}
\newblock \bibinfo{title}{Crystal-like topological defect arrays in nematic
  liquid crystal}.
\newblock \emph{\bibinfo{journal}{Applied Physics Letters}}
  \textbf{\bibinfo{volume}{119}} (\bibinfo{year}{2021}).

\bibitem{subhash2011full}
\bibinfo{author}{Subhash, H.~M.} \& \bibinfo{author}{Rosen, J.}
\newblock \bibinfo{title}{Full-field and single-shot full-field optical
  coherence tomography: a novel technique for biomedical imaging applications}.
\newblock \emph{\bibinfo{journal}{Advances in Optical Technologies}}
  \textbf{\bibinfo{volume}{2012}}, \bibinfo{pages}{85} (\bibinfo{year}{2011}).

\bibitem{RTweezerPiestun}
\bibinfo{author}{Conkey, D.}, \bibinfo{author}{Trivedi, R.},
  \bibinfo{author}{Pavani, S.}, \bibinfo{author}{Smalyukh, I.} \&
  \bibinfo{author}{Piestun, R.}
\newblock \bibinfo{title}{Three-dimensional parallel particle manipulation and
  tracking by integrating holographic optical tweezers and engineered point
  spread functions}.
\newblock \emph{\bibinfo{journal}{Optics Express}}
  \textbf{\bibinfo{volume}{19}}, \bibinfo{pages}{3835--3842}
  (\bibinfo{year}{2011}).

\bibitem{martinez2014mutually}
\bibinfo{author}{Martinez, A.} \emph{et~al.}
\newblock \bibinfo{title}{Mutually tangled colloidal knots and induced defect
  loops in nematic fields}.
\newblock \emph{\bibinfo{journal}{Nature materials}}
  \textbf{\bibinfo{volume}{13}}, \bibinfo{pages}{258--263}
  (\bibinfo{year}{2014}).

\bibitem{smalyukh2010three}
\bibinfo{author}{Smalyukh, I.~I.}, \bibinfo{author}{Lansac, Y.},
  \bibinfo{author}{Clark, N.~A.} \& \bibinfo{author}{Trivedi, R.~P.}
\newblock \bibinfo{title}{Three-dimensional structure and multistable optical
  switching of triple-twisted particle-like excitations in anisotropic fluids}.
\newblock \emph{\bibinfo{journal}{Nature materials}}
  \textbf{\bibinfo{volume}{9}}, \bibinfo{pages}{139--145}
  (\bibinfo{year}{2010}).

\bibitem{kos2022nematic}
\bibinfo{author}{Kos, {\v{Z}}.} \& \bibinfo{author}{Dunkel, J.}
\newblock \bibinfo{title}{Nematic bits and universal logic gates}.
\newblock \emph{\bibinfo{journal}{Science advances}}
  \textbf{\bibinfo{volume}{8}}, \bibinfo{pages}{eabp8371}
  (\bibinfo{year}{2022}).

\end{thebibliography}

\newpage
\end{document}